\documentclass[graphicx,floatfix,aps,pre,epsfig,showpacs]{revtex4}
\input{psfig.sty}
\usepackage{graphicx}
\usepackage{amsmath}
\begin{document}

\title{Message passing for vertex covers}

\author{Martin Weigt}
\affiliation{Institute for Scientific Interchange, Viale Settimio Severo 65,
I-10133 Torino, Italy}

\author{Haijun Zhou}
\affiliation{Institute of Theoretical Physics, The Chinese
Academy of Sciences, Beijing 100080, China}

\date{\today}

\begin{abstract}
Constructing a minimal vertex cover of a graph can be seen as a
prototype for a combinatorial optimization problem under hard
constraints. In this paper, we develop and analyze message passing
techniques, namely warning and survey propagation, which serve as
efficient heuristic algorithms for solving these computational hard
problems. We show also, how previously obtained results on the
typical-case behavior of vertex covers of random graphs can be
recovered starting from the message passing equations, and how they
can be extended.
\end{abstract}

\pacs{89.75.-k Complex systems, 02.10.Ox Combinatorics, graph theory,
75.10.Nr Spin-glass and other random models}

\maketitle 


\section{Introduction}

The minimal vertex-cover (VC) problem belongs to the most difficult
class of optimization problems in graph theory \cite{GareyM1979}. It
asks to mark a minimal number of vertices of a graph, such that each
edge of the graph is incident to at least one of the selected
vertices. This problem is known to be NP-hard, which means in
particular that all currently known algorithms construct minimal
vertex covers in a computational time which scales exponentially with
the size of the graph. The applicability of such exact algorithms is
therefore restricted to pretty small graphs of few hundreds of
vertices.

There are, however, applications of the vertex covering problem and
other, closely related optimization problems \cite{VaWe,Va} to a huge
number of real-world network problems, examples are the monitoring of
Internet traffic \cite{Breitbart}, the prevention of denial-of-service
attacks \cite{Park} and immunization strategies in networks
\cite{Gomez}. Another technically related network problem is the one
of counting loops in networks, recently analyzed on the basis of
statistical-physics methods \cite{SeMaMo}.  The dimensions of the
underlying networks easily exceed the graph sizes treatable with exact
algorithms, and heuristic methods to construct as small as possible
solutions are needed.

In this paper we set up two message-passing techniques based on the
statistical-physics approach to combinatorial optimization problems
\cite{HaWeBook,Percus}, more precisely based on the cavity method for
diluted systems \cite{MezardM2001} and its algorithmic interpretation
\cite{YeFrWe,MezardM2002}. The first of the message
passing-techniques, the so-called warning-propagation, is equivalent
to the Bethe-Peierls iterative scheme and therefore related to the
assumption of replica symmetry, the second one is a survey propagation
algorithm related to one-step replica symmetry breaking. Both
algorithms have already been formulated for the vertex cover problem
by one of the authors in \cite{HaWeBook}, here we go beyond this
presentation providing both a more elegant setting and a thorough
analysis of the algorithmic performance.

A natural test bed for the proposed algorithms is provided by
finite-connectivity random graphs \cite{BollobasB1985}. The typical
properties of VCs on such graphs have already been analyzed both with
rigorous mathematical tools \cite{Gaz,Fr} and with statistical-physics
methods \cite{WeigtM2000, BauerM2001, WeigtM2001,ZhouH2003,ZhouH2005a}. 
We therefore have a pretty complete knowledge
about the phase diagram of this problem, and can systematically compare the
algorithmic performance of the message-passing techniques on single,
finite, randomly generated graphs to the average behavior in the
thermodynamic limit.

It should also be noted that the vertex cover problem is closely
related to a class of lattice glass models
\cite{BiMe,HaWe,Co,RiBiMeMa,HanWe}. In these models, hard particles
are to be positioned on a lattice under geometrical packing
constraints representing hard-core interactions. These models are
considered as simple lattice models for the glass transition due to
geometric frustration, and their closest packing correspond to minimal
VCs \cite{WeigtM2001}.

This paper is organized as follows: the vertex cover problem is
defined in Sec.~\ref{sec:model} and the concept of cavity graph is
defined in Sec.~\ref{sec:cavity}; Sec.~\ref{sec:wp} focuses on the
warning propagation algorithm, and its performance and iterative
stability are analyzed; Sec.~\ref{sec:sp} focuses on the survey
propagation algorithm; and Sec.~\ref{sec:mf} estimates the minimal
vertex cover size for a random graph using statistical physics method;
finally in Sec.~\ref{sec:conclusion} we conclude this work.

\section{The model}
\label{sec:model}

Let us start with the definition of vertex covers. Given is a graph
$G=(V,E)$ with $N$ vertices $i=1,...,N$ and $M$ undirected edges
$\{i,j\}=\{j,i\}\in E$ connecting pairs of vertices.\\

{\bf Definitions:} A {\it vertex cover} (VC) of the graph $G$ is a
subset $U\subset V$ of vertices such that for all edges $\{i,j\}\in E$,
at least one end vertex is element of $U$, i.e. $i\in U$ or $j\in U$.
A {\it minimal vertex cover} of $G$ is a vertex cover of minimal
cardinality.\\

We also denote vertices in $U$ as covered, as well as their incident
edges: The set $U$ is a VC iff all edges are covered. Determining a
minimal vertex cover is one of the basic NP-hard combinatorial
problems \cite{GareyM1979}. Its worst-case solution time is consequently
expected to grow exponentially with the size of the problem instance,
here measured by the vertex and edge numbers $N$ and $M$. The problem
is equivalent to the problem of constructing a {\it maximum
independent set} of $G$, and to the problem of finding the {\it
maximum clique} in the complementary graph of $G$ (where edges and
non-edges are exchanged).

The exponential running time of algorithms constructing minimal vertex
covers is a serious limitation to practical applications: Exact
algorithms are able to treat only relatively small sample graphs. It is
therefore interesting to develop powerful heuristic methods which are
able to construct at least close-to-minimal VCs, which may serve as
reasonable solutions in practical problems.

In  the context  of constraint-satisfaction  problems  (CSP), recently
statistical-physics approaches  have led to the  proposal of so-called
{\it   survey   propagation   algorithms}  which   are   sophisticated
message-passing procedures  based on the cavity  method of statistical
physics.  This   type  of  algorithm   was  first  proposed   for  the
satisfiability  problem  \cite{MezardM2002},   and  then  extended  to
graph-coloring      \cite{PagnaniA2003}      and     general      CSPs
\cite{BraunsteinA2005}, and is one of the most efficient algorithms in
the hard-to-solve phase of these problems.

The vertex cover problem is structurally different from CSPs. Whereas
the computational problem of the latter results from the existence of
a large number of constraints being hard to satisfy simultaneously,
the constraints in vertex cover -- i.e.~the need of covering each edge
of the graph -- can in principle be satisfied very easily by covering
many vertices. The computational hardness stems from the objective of
finding a {\it minimal} vertex cover, i.e.~from the interaction between a
high number of local constraints on one side, and the global
minimization condition on the other side. This leads to a difference
in the validation of the output of a heuristic algorithm: Whereas a
solution to a CSP can be easily checked by testing all constraints,
and the problem consists in finding one, it is no problem at all to
construct a VC, but its minimality can hardly be shown. One can say
that the hardness of solving VC stems from the fact that the landscape
$X(U)=|U|$ becomes complex over the set of all VCs $U$ (note: not over
the set of all vertex sub-sets).

The algorithmic aim is therefore to construct a vertex cover as small
as possible in polynomial time for some given graph $G=(V,E)$. The
central step in this context will be the calculation (or at least
approximation) of the vertex-dependent number
\begin{equation}
  \label{eq:sp_pi}
  \pi_i = \frac{ |\{ U\subset V\ |\ U {\mbox{ is min. VC}},\ i\in
  U\}|}{ |\{ U\subset V\ |\ U {\mbox{ is min. VC}} \}|}
\end{equation}
which, for every vertex $i\in V$, equals the fraction of minimal vertex
covers containing vertex $i\in V$. In probabilistic terms, it can be
understood as the probability that $i$ is covered in a randomly
selected minimal vertex cover.

Once we know these quantities, we can obviously exploit them
algorithmically. We know, e.g., that each vertex with $\pi_i = 1$
belongs to all minimal VCs, and it has to be included into the VC we
are aiming to construct. Contrarily, vertices $i\in V$ with $\pi_i =
0$ do not appear in any minimal VC, and they have to be excluded from
the vertex set we are building. The problem is slightly more involved
for those vertices having $\pi$-values different from zero and one:
They are contained in some vertex covers, but not in others. Since
$\pi_i$ gives only a strictly local information, we do not know any
possible quantitative restriction to the simultaneous assignment of
pairs or even larger subsets of vertices. If we consider, e.g., one
edge $\{i,j\}\in E$, the joint probability that both vertices are
uncovered does not equal $(1-\pi_i)(1-\pi_j)$ as one might assume
naively by considering the vertices to be independent.  It equals
obviously zero due to the vertex-cover constraint for the edge: At
least one of the end-vertices has to be covered. This problem can be
resolved by an iterative decimation process. We select, e.g., a vertex
of non-zero $\pi$ and add it to the VC $U$ to be constructed, and
delete the vertex as well as all its incident edges from the graph. We
than recompute the $\pi$ from the decimated graph, add a new vertex to
$U$ and so on, until all edges of $G$ are covered: The vertex set $U$
now forms a vertex cover of the graph $G$.

There is an obvious algorithmic problem with evaluating the $\pi_i$: A
naive calculation according to their definition would require the
prior knowledge of all minimal VCs - which we do not have if we are
trying to develop an algorithm finding just a single one of them. The
way out will be a message passing procedure \cite{YeFrWe,Kc} which
only exchanges local information between neighboring vertex pairs,
until these messages reach globally self-consistent values. Such
message passing procedures first need the introduction of the {\it
cavity graph}, which will be done in the following section.

\section{The cavity graph}
\label{sec:cavity}

A simple idea could be to determine $\pi_i$ from all the $\pi_j$ of
the neighbors $j \in N(i)$ of vertex $i$. This is not directly
possible: As discussed above, the $\pi_j$ are single-site quantities
and do not contain any information of vertex pairs. Any two $j\in
N(i)$ are, however -- via a path crossing $i$ -- second neighbors of
each other, and thus they are highly correlated. Imagine, e.g., that
vertex $i$ is not covered, than {\it all} $j\in N(i)$ have to be
simultaneously covered. The knowledge of the marginal cover
probabilities $\pi_j$ is obviously not sufficient to determine also
the central $\pi_i$.  The way out is to consider not the full graph,
but the {\it cavity graphs}:\\

{\bf Definition:} Given a graph $G=(V,E)$, and a vertex $i\in V$, the
{\it cavity graph} $G_i$ is the subgraph of $G$ induced by the vertex
set $V_i = V \setminus i$.\\

Said with simpler words, the cavity graph is created from the full
graph $G$ by removing vertex $i$ as well as its incident edges
$\{i,j\}$ for all $j\in N(i)$. On a tree graph, the $j\in N(i)$ would
belong to pairwise distinct connected components of the cavity graph,
and they would be independent of each other. More generally, on a
graph with relatively long cycles, any two of the former neighbors of
vertex $i$ will be distant on the cavity graph $G_i$. The basic
approximation underlying message passing algorithms consists in
assuming statistical independence of these vertices on the cavity
graph (within one thermodynamic state, as will be explained in the
case of survey propagation).

Having defined the cavity graphs $G_i$ for each vertex $i$, we also
define the generalized probabilities
\begin{equation}
  \label{eq:sp_pi_cav}
  \pi_{j|i} = \frac{ |\{ U\subset V_i\ |\ U {\mbox{ is min. VC of }}
  G_i,\ j\in U\}|}{ |\{ U\subset V_i\ |\ U {\mbox{ is min. VC of }} G_i
  \}|}
\end{equation}
measuring the fraction of minimal vertex covers of the cavity graph
$G_i$ containing vertex $j\neq i$. Even if defined formally for any
pair of vertices $i$ and $j$, these quantities will be relevant in
particular for those vertices connected by an edge in the original
graph $G$, i.e.~for $\{i,j\}\in E$.

A comment on the statistical-independence assumption has to be
included at this point: We are constructing an algorithm for real,
i.e. finite graphs. This means that graph loops have finite
length. The equations we are going to present in the following will
therefore be only approximations to the exact values of the
probabilities $\pi_i$, and the algorithm cannot guarantee to construct
a true minimum vertex cover. So, even if the presented algorithm will
scale only quadratically in the graph order $N$, it cannot be
considered as an exact polynomial algorithm, and therefore does not
contribute to the solution of the P-NP problem. The importance of
message passing algorithms is related to practical applications on
large graphs, where exact methods fail due to their exponential time
requirements. As we will see below in numerical simulations, the
procedures presented here largely outperform purely local algorithms,
and therefore allow to construct better approximations to the exact
solution.

\section{Warning propagation (WP)}
\label{sec:wp}

\subsection{The algorithm}
\label{sec:wp_alg}

The very first and simplest message passing procedure we are going to
introduce, carries the name {\it warning propagation} (WP). In this
case, we are going to calculate only the reduced quantities
\begin{equation}
  \label{eq:pi_tilde}
  \tilde\pi_i = \left\{
  \begin{array}{ll}
    0 & {\rm if}\ \pi_i=0\\
    * & {\rm if}\ 0 < \pi_i < 1\\
    1 & {\rm if}\ \pi_i=1 
  \end{array}
  \right.
\end{equation}
and analogously the cavity quantities $\tilde\pi_{j|i}$. So these
quantities are not measuring the exact probability for a vertex to be
covered in a randomly selected minimal vertex cover. They only
indicate whether it is always covered (value one), never covered
(value zero) or sometimes covered and sometimes uncovered. For this
last case we have introduced the unifying {\it joker state} $*$. Note
that also this information is sufficient to be exploited
algorithmically: If a vertex is assigned the joker state, it can be
chosen liberally to be covered or to be uncovered during graph
decimation.

As a first step, we introduce an even simpler message type, the
so-called warning $u_{j\to i}$ sent from a vertex $j$ to a neighbor
$i$. This warning incorporates the vertex cover constraint: If the
vertex $j$ is uncovered, it sends a warning $u_{j\to i}=1$ to vertex
$i$ signifying: ``Attention, to cover our connecting edge you should
be covered, or I have to change state.''  If, on the other hand,
vertex $j$ is already covered, it sends the trivial message $u_{j\to
i}=0$ saying: ``I have already covered our connecting edge.''  More
formally, a set of warnings is defined for every vertex subset
$U\subset V$:
\begin{equation}
  \label{eq:warning}
  u_{j\to i}(U) := \left\{
  \begin{array}{ll}
    0 & {\rm if}\ j\in U\\
    1 & {\rm if}\ j\notin U 
  \end{array}
  \right.
\end{equation}
with $\{i,j\}\in E$ being an arbitrary edge. Note that each edge
carries {\it two} messages: One sent from $i$ to $j$, the other one
from $j$ to $i$. In a proper VC, at least one of the end-vertices of
each edge has to be covered, so we find that
\begin{equation}
  \label{eq:warning_vc}
  U\subset V {\mbox{ is VC of }} G\ \ \ \ \leftrightarrow \ \ \ \ 
  \forall \{i,j\}\in E:\
  u_{i\to j}(U)\cdot u_{j\to i}(U)=0\ ,
\end{equation}
i.e. each edge has to carry at least one trivial warning. The
definition of the warning can also be extended to sets ${\cal M}$ of
vertex subsets. We define
\begin{equation}
  \label{eq:warning_set}
  u_{j\to i}({\cal M}) := \min_{U\in{\cal M}} \ u_{j\to i}(U) \ ,
\end{equation}
i.e. a non-trivial message is sent if and only if vertex $j$ is
element of no $U\in{\cal M}$. This definition obviously reproduces the
warning (\ref{eq:warning}) if ${\cal M}$ consists of only one vertex
subset. The reason for selecting the minimum in the last definition
will become clear below. Using the set ${\cal S}_i$ of all minimal
vertex covers of the cavity graph $G_i$ as a special case, the warning
$ u_{j\to i}({\cal S}_i)$ becomes a function of $\tilde\pi_{j|i}$
only. For an arbitrary but fixed edge we find
\begin{equation}
  \label{eq:warning_sols}
  u_{j\to i}({\cal S}_i) \equiv  u_{j\to i}(\tilde\pi_{j|i}) =
  \left\{
  \begin{array}{ll}
    1 & {\rm if}\ \tilde\pi_{j|i}=0\\
    0 & {\rm if}\  \tilde\pi_{j|i}=*\\
    0 & {\rm if}\  \tilde\pi_{j|i}=1
  \end{array}
  \right. \ .
\end{equation}

\begin{figure}[htb]
  \vspace{0.2cm}
  \begin{center}
    \includegraphics[width=0.3\columnwidth]{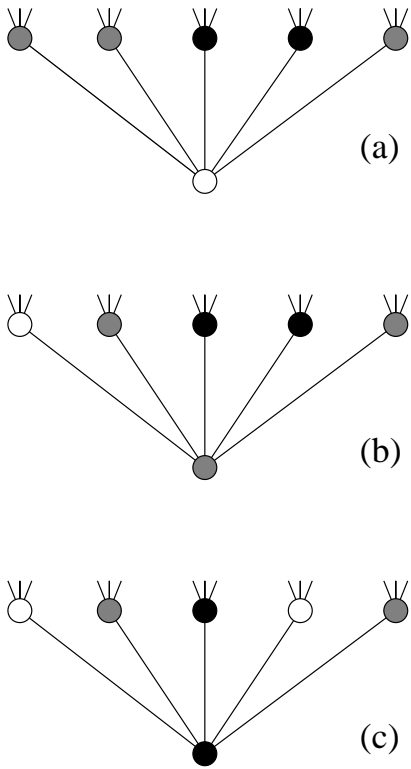}
  \end{center}
  \caption{Graphical representation of Eq.~(\ref{eq:pi_iter}), with
    vertex $i$ being identified with the lower vertex in each
    sub-figure. The color coding of the vertices corresponds to the
    values of $\tilde\pi_i$ and the $\tilde\pi_{j|i}$: Value zero is
    represented by a white dot, value one by a black dot, and the
    joker value $*$ by a gray dot. In case (a), there are no white
    dots between the $j\in N(i)$, so the lower vertex has not to be
    covered and gets color white. If there is exactly one white dot in
    the upper line, the lower vertex becomes gray, cf. (b). If there
    are two or more white dots in the upper line, as in (c), the lower
    vertex is black, corresponding to an always covered vertex.}
  \label{fig:messages}
\end{figure}

The required minimality of the vertex cover to be constructed leads to
a simple propagation of these warnings, or equivalently of the
corresponding $\tilde\pi_{j|i}$. This can be achieved by considering
how minimal vertex covers can be extended from the cavity graph to the
full graph. There are three cases, cf.~Fig.~\ref{fig:messages}:
\begin{enumerate}
\item[{ (a)}] There exists at least one minimal vertex cover of the
  cavity graph $G_i$ where {\it all} $j\in N(i)$ (neighbors of $i$ in
  the full graph $G$) are simultaneously covered. These VCs are also
  minimal VCs of the full graph $G$ since all edges incident to $i$
  are already covered, so $i$ has to be uncovered to guarantee
  minimality. The sizes of the minimal VCs of $G_i$ and those of $G$
  thus coincide. In this case we find $\tilde\pi_i=0$, since there are
  no minimal VCs of $G$ containing $i$.
\item[{ (b)}] All minimal vertex covers of $G_i$ leave at least
  two $j\in N(i)$ uncovered. Since all edges incident to vertex $i$
  have to be covered, we have to add $i$ to the VC of $G_i$ in order
  to extend it to the full graph. The VC of the full graph contains
  thus exactly one vertex more than those of the cavity graph, and
  $\tilde\pi_i$ equals one.
\item[{ (c)}] In the last, intermediate case, there is at least
  one minimal VC of $G_i$ containing all but one $j\in N(i)$, but
  there is none containing all $j\in N(i)$. Also in this case, we have
  to add exactly one vertex by going from a VC of the cavity graph
  $G_i$ to one of the full graph $G$, the VC size grows by one. If we,
  however, use the VC leaving only one $j\in N(i)$ uncovered, there
  exists only one single uncovered edge in $G$. To cover it, we can
  select any one of its two end vertices, i.e. either $i$ or its
  neighbor. In this case, we therefore find $\tilde\pi_i=*$,
  i.e. vertex $i$ is found to be in the joker state.
\end{enumerate}
At this point, the independence assumption of all $j\in N(i)$ in the
cavity graph enters into the discussion: We consider their joint
probability of simultaneously being covered in a minimal VC of the
cavity graph $G_i$, and assume this quantity to factorize into
$\prod_{j\in N(i)} \pi_{j|i}$. Under this assumption, case (a)
happens if and only if all 
$\tilde\pi_{j|i}\neq 0$.
Case (b) happens if there
are at least two vanishing $\tilde\pi_{j|i}$ in between the $j\in
N(i)$, and the third case appears for exactly one zero
$\tilde\pi_{j|i}$. We see that in this rule no difference between
always covered and joker vertices $j\in N(i)$ exists, which explains
the use of the minimum warning in definition~(\ref{eq:warning_set}). We
conclude:
\begin{equation}
  \label{eq:pi_iter}
  \tilde\pi_i =
  \left\{
  \begin{array}{ll}
    0 & {\rm if}\ \sum_{j\in N(i)} u_{j\to i}(\tilde\pi_{j|i}) = 0\\
    * & {\rm if}\ \sum_{j\in N(i)} u_{j\to i}(\tilde\pi_{j|i}) = 1\\
    1 & {\rm if}\ \sum_{j\in N(i)} u_{j\to i}(\tilde\pi_{j|i}) > 1
  \end{array}
  \right. \ .
\end{equation}
This rule is graphically represented in Fig.~\ref{fig:messages}.
The cavity quantities $\tilde\pi_{j|i}$ can now be calculated by
considering the cavity graphs $G_j$, and by disregarding in addition
the influence of vertex $i$:
\begin{equation}
  \label{eq:pi_cav_iter}
  \tilde\pi_{j|i} =
  \left\{
  \begin{array}{ll}
    0 & {\rm if}\ \sum_{k\in N(j)\setminus i} 
    u_{k\to j}(\tilde\pi_{k|j}) = 0\\
    * & {\rm if}\ \sum_{k\in N(j)\setminus i} 
    u_{k\to j}(\tilde\pi_{k|j}) = 1\\
    1 & {\rm if}\ \sum_{k\in N(j)\setminus i} 
    u_{k\to j}(\tilde\pi_{k|j}) > 1
  \end{array}
  \right. \ .
\end{equation}
Eqs.~(\ref{eq:warning_sols}-\ref{eq:pi_cav_iter}) are called {\it
warning propagation}. The last equation, together with
Eq.~(\ref{eq:warning_sols}), describes a closed system of $2|E|$
equations: Two for each edge due to the two different possible
orientations of the messages. Note that Eqs.~(\ref{eq:warning_sols})
and (\ref{eq:pi_cav_iter}) can also be reformulated for the warnings
$u_{j\to i}$ itself, eliminating the cavity quantities
$\tilde\pi_{j|i}$. The iterative equations take the particularly
simple form
\begin{equation}
u_{j\to i} = \delta \left( \sum_{k\in N(j)\setminus i} u_{k\to j}, 0
    \right)
\label{eq:u_iter}\ ,
\end{equation}
where, for better readability, we have used the notation
$\delta(\cdot,\cdot)$ for the Kronecker symbol.

These equations have to be solved and plugged into
Eq.~(\ref{eq:pi_iter}) in order to calculate the values of all
$\tilde\pi_i$. Even if this information is not yet sufficient to
immediately solve the minimal VC problem, we can already read off a
lot of useful information about the properties of all minimal vertex
covers. The most important quantity is an estimate for the minimal VC
cardinality:
\begin{equation}
\label{eq:sizevcwp}
X = \sum_{i\in V} \left[ \delta(\tilde\pi_i,1)+\frac 12
\delta(\tilde\pi_i,*)  \right]\ .
\end{equation}
The prefactor 1/2 in front of the number of joker vertices is not a
direct result of WP. It can be justified using the more detailed
belief propagation calculating the full single site probability
$\pi_i$ \cite{HaWeBook}, or via the replica method \cite{WeigtM2001}.

The WP equations can be used to construct a vertex cover, i.e. they
can be exploited algorithmically. This is done in the following way,
starting with an initial graph $G=(V,E)$ and an empty set
$U=\emptyset$:
\begin{enumerate}
\item The $2|E|$ warnings $u_{i\to j}$ are initialized randomly. 
\item Then, sequentially, edges are selected and the warnings are
  updated using Eq.~(\ref{eq:u_iter}). This update is iterated until a
  solution of the warning-propagation equations is found.
\item The $\tilde\pi_i$ are calculated from the warnings using
  Eq.~(\ref{eq:pi_iter}).
\item All vertices with $\tilde\pi_i=1$ are added to $U$, and deleted
  with their incident edges from $G$.
\item All vertices with $\tilde\pi_i=0$ are deleted from $V$, without
  changing $U$. Since a vertex with $\tilde\pi_i=0$ has only neighbors
  of $\tilde\pi_i=1$, it was already isolated after the last step. No
  edges have therefore to be removed from $E$.
\item One remaining vertex $i$ ($\tilde\pi_i=*$) is selected, and all
  its neighbors $j\in N(i)$ are added to $U$. Vertices $i$ and $N(i)$
  are removed from $V$, and all their incident edges are subtracted
  from $E$.
\item If uncovered edges are left, we go back to step 2, and
  recalculate the warnings on the decimated graph. If no edges are
  left, the current $U$ is returned as a solution.
\end{enumerate}
Obviously, the constructed $U$ forms a vertex cover, since only
covered edges are removed from the graph. It is also a minimal one, if
the information provided by the $\tilde\pi_i$ was correct. Due to the
factorization hypothesis in WP, some of the $\tilde\pi_i$ may,
however, be erroneous, resulting possibly in a non-minimal cover. It
is worth to note that after each graph decimation step followed by a
re-iteration of the WP equations, a new estimate of the VC size can be
calculated according to Eq.~(\ref{eq:sizevcwp}). This estimate is
expected to be stationary only in the case where already the initial
warnings where exact, and to change under the algorithm if the latter
were only approximations.

\subsection{From single samples to average results on random graphs}
\label{sec:wp_to_rs}

Starting from Eqs.~(\ref{eq:u_iter}) and (\ref{eq:pi_iter}), we can
easily reconstruct the replica-symmetric typical-case results for
random graphs of average vertex degree $c$. We start with defining the
global histogram of warnings,
\begin{equation}
Q(u) = \frac 1{2|E|} \sum_{(i,j)\in E} \left[ \delta(u_{i\to j},u) +
\delta(u_{j\to i},u) \right]\ .
\label{eq:u_hist}
\end{equation}
Due to the binary nature of the warnings, it can be parametrized as
\begin{equation}
Q(u) = \rho_0 \delta(u,0) + \rho_1 \delta(u,1)
\end{equation}
with $\rho_0+\rho_1=1$. Consider now Eq.~(\ref{eq:u_iter}): A
non-trivial warning is sent via a link $j\to i$ only if the input
messages $u_{k\to j}$ from all $k\in N(j)\setminus i$ equal zero. This
happens for all warnings independently with probability $\rho_0$, and
the number $d$ of these incoming messages is, on a random graph,
distributed according to a Poissonian of mean $c$. We thus find
\begin{equation}
\rho_1 = \sum_{d=0}^\infty e^{-c} \frac{c^d}{d!} \rho_0^d 
= e^{-c \rho_1}\ ,
\end{equation}
which, using the Lambert-$W$ function, is solved by
\begin{equation}
\rho_1 =  \frac{W(c)}{c}\ .
\label{eq:rho1}
\end{equation}

Let us now reconstruct also the histogram
\begin{equation}
P(\tilde\pi) = \frac 1N \sum_{i\in V} \delta(\tilde\pi_i, \tilde\pi)
\end{equation}
of single-site marginals $\tilde\pi_i$. The latter are three-valued,
we thus parametrize the histogram as
\begin{equation}
P(\tilde\pi) = \nu_0 \delta(\tilde\pi,0) + \nu_* \delta(\tilde\pi,*) 
+ \nu_1 \delta(\tilde\pi,1)\ .
\end{equation}
Using Eq.~(\ref{eq:pi_iter}) and the Poissonian degree distribution,
we find
\begin{eqnarray}
\nu_0 &=& \sum_{d=0}^\infty e^{-c} \frac{c^d}{d!} \rho_0^d \nonumber\\
&=& \rho_1 \nonumber\\ 
\nu_* &=& \sum_{d=0}^\infty e^{-c} \frac{c^d}{d!} 
d \rho_0^{d-1} \rho_1 \nonumber\\ 
&=& c \rho_1^2 \nonumber\\
\nu_1 &=& 1-\nu_*-\nu_0
\end{eqnarray}
For the derivation of the second expression we have used that the single
non-zero warning among the messages reaching a joker vertex can be
chosen liberally in between all $d$ incoming edges. For the VC size
we thus find
\begin{equation}
x(c) = \lim_{N\to\infty} \frac XN = 1-\frac{W(c)}{c} - \frac{W(c)^2}{2c}
\end{equation}
which is identical to the result of a replica-symmetric calculation
\cite{WeigtM2000}. For average degrees $c<e$, this result was shown to
be exact \cite{BauerM2001}, and it can in fact already be read off
from an older result by Karp and Sipser \cite{KaSi} on maximal
matchings in random graphs, see also \cite{OuZh,ZdMe} for related
statistical-physics approaches.

\subsection{Bug proliferation and the stability of the WP fixed point}
\label{sec:wp_stab}

Besides the problem that the solution of the equations of warning
propagation may be imprecise due to the existence of short loops in
the graph, there can be another problem - the iteration of the warning
update may fail to converge. This can happen again due to the
existence of short loops, which may lead to attractive limit cycles in
the iterative warning dynamics. Another problem can appear due to the
existence of {\it many} solutions of Eqs.~(\ref{eq:pi_cav_iter}). In
statistical physics we say that the replica symmetry is broken.

To be more quantitative, we study here the stability of a WP solution
with respect to the introduction of a {\it bug} \cite{MeMeZe}:
One of the warnings $u_{j\to i}$ is changed to its opposite
value. After one iteration of WP, the bug itself will be cured since
it depends only on unchanged messages. On the other hand, the warnings
from vertex $i$ to its neighbors $k\in N(i)\setminus j$ may be
changed, i.e.~new bugs may appear. The question is now if these bugs
proliferate and, after some iterations, change a finite fraction of
all warnings, or, if the bugs die out after a while. Only in the
second case, WP is stable and can be usefully included into a
decimation procedure.

Here, we perform this analysis analytically for the case of a random
graph of average degree $c$. In this case, the number $d$ of neighbors
$k\in N(i)\setminus j$ receiving messages depending on the bug is
distributed according to the Poissonian $e^{-c} c^d/d!$. They send
themselves warnings $u_{k\to i}$ to vertex $i$ which are, due to the
locally tree-like structure of a random graph, independent on $u_{j\to
i}$, and can be considered to be randomly selected according to the
global histogram $Q(u)=\rho_0 \delta(u,0) + \rho_1 \delta(u,1)$ of
warnings introduced in Eq.~(\ref{eq:u_hist}).

We have to distinguish two cases for the introduction of a bug:
\begin{enumerate}
\item[{\it (i)}] We change the message $u_{j\to i}$ from {\it one to zero}.

  Prior to this change, all out-messages $u_{i\to k}$ with $k\in
  N(i)\setminus j$ were equal to zero, cf.~Eq.~(\ref{eq:u_iter}). Let
  us denote by $d=|N(i)\setminus j|$ the number of the out-messages
  depending on $u_{j\to i}$, i.e., the degree of vertex $i$ equals
  $d+1$.

  After the introduction of the bug, an out-message $u_{i\to k}$ becomes
  one if and only if all other in-messages $u_{l\to i}$ with $l\in
  N(i)\setminus \{j,k\}$ are zero. There are two sub-cases. First,
  with probability $\rho_0^d$, all messages $u_{k\to i}$ equal
  zero. In this case, all $d$ out-messages change. Second, with
  probability $d \rho_0^{d-1} (1-\rho_0)$, exactly one message
  $u_{k\to i}$ has value one, all other zero. In this case, only
  $u_{i\to k}$ changes under iteration. On average, the bug introduces 
  thus
  $$ 
  \sum_d e^{-c} \frac{c^d}{d!} \left[ d \rho_0^d + d \rho_0^{d-1}
  (1-\rho_0) \right] = c\ e^{-c(1-\rho_0)} = c\ e^{-c \rho_1}
  $$
  new bugs into the graph. These bugs are of the second type.
\item[{\it (ii)}] We change the message $u_{j\to i}$ from {\it zero to one}.

  After introduction of the bug, all out-messages $u_{i\to k}$ with
  $k\in N(i)\setminus j$ become zero under WP update. They are bugs
  only if, in the initial WP solution, they had the value one.
  Using analogous arguments to the first case, we find that, with
  probability $\rho_0^d$, all $d$ out-messages were one, and, with
  probability $d \rho_0^{d-1} (1-\rho_0)$, only one single message was
  one and becomes a new bug. The expected number of new bugs caused
  by the changed $u_{j\to i}$ equals again $c\ e^{-c \rho_1}$.
\end{enumerate}
We now apply a simple percolation-type argument: If the average number
of new bugs is smaller than one, the bugs are expected to be cured
after a few iterations, the WP solution is stable under bug
introduction. If, on the other hand, the average number of new bugs is
larger than one, we expect an exponential increase in the bug
number. Bugs proliferate and carry away the system from the WP fixed
point. The latter is thus concluded to be unstable. Note that this
arguments holds only because we update out-messages which, under
iterated WP updates, do not interact because they all influence
disjoint sets of further warnings.

The critical point can now be determined easily: The average number of
new bugs is set to one, $c_{WP}\ e^{-c_{WP} \rho_1}=1$. Comparing it to
the self-consistent Eq.~(\ref{eq:rho1}), we immediately conclude the
\begin{equation}
c_{WP} = \frac 1 {\rho_1(c_{WP})} = e\ .
\end{equation} 
WP converges below this critical connectivity, i.e.~in the full region
where replica symmetry is exact. As one would expect intuitively, it
does not converge in the replica-symmetry broken phase above average
degree $e$, there survey propagation as discussed in the following
section of this work has to be applied.

\subsection{Bug relaxation time of WP}
\label{sec:wp_time}

Even if WP provides asymptotically exact results in the full replica
symmetric phase in a running time scaling quadratically with $N$, its
convergence slows down if we approach the critical average
degree. This can be seen analytically by calculating the evolution of
the number of erroneous messages, or bugs, under various update
schemes.

\subsubsection{Parallel update}

Let us start with a parallel update scheme, where, in every iteration
step, all messages are recalculated simultaneously from the old
messages. Assume, that there are $M_1 \ll 2M = 2 |E|$ erroneous
messages. These are, up to higher-order effects, isolated from each
other and act thus independently under WP iteration. Each of these
bugs becomes thus corrected in a new WP step, but causes, as seen in
the last sub-section, on average $c\ e^{-c \rho_1} = c\rho_1$ new
wrong messages. Again, up to higher order corrections, these messages
do not interact. For the expected number $\overline{M_1(t)}$ we thus
find
\begin{equation}
\overline{M_1(t)} = (c\rho_1)^t \overline{M_1(0)}\ ,
\end{equation}
i.e., below $c_{WP}=e$, this number decays exponentially with a time
scale
\begin{equation}
\tau_{par} = - \frac 1 {\ln (c\rho_1)}\ .
\end{equation}
This relaxation time diverges is we approach $c=e$ from below. To
unveil the critical behavior, we set $c=e-\varepsilon$
($0<\varepsilon\ll 1$). With $\rho_1 = 1/e +\delta$ we find, using
Eq.~(\ref{eq:rho1}),
\begin{equation}
\frac 1e + \delta = \exp\left\{-1+\frac \varepsilon e - e \delta +
{\cal O}(\varepsilon\delta) \right\}\ ,
\end{equation}
i.e. $\delta=\varepsilon/(2e^2)$, resulting in $c\rho_1 =
1-\varepsilon/(2e)+{\cal O}(\varepsilon^2)$, and thus in
\begin{equation}
\tau_{par} \simeq \frac{2e}{e-c} \ \ \ \ \ \ \ \ 
{\rm for}\ \ 0<e-c\ll 1 \ .
\end{equation}
The critical exponent one is expected to result from the mean-field
structure of the underlying graph.

\subsubsection{Random update}

The situation is slightly more involved in the case of a random
update, where in every time step one message is selected randomly out
of all $2M$ warnings, and is updated according to the WP equation. Let
us denote by $p_T(M_1)$ the probability that there are $M_1$ erroneous
messages after $T$ random update steps. Its evolution under WP is
given by the rate equation
\begin{equation}
p_{T+1} (M_1) = p_T (M_1)-\frac {M_1}{2M} p_T(M_1) 
+ \frac{M_1+1}{2M} p_T(M_1+1) - \frac{c\rho_1 M_1}{2M} p_T(M_1)
+ \frac{c\rho_1 (M_1-1)}{2M} p_T(M_1-1)\ .
\label{eq:rate_update}
\end{equation}
This is due to the fact, that, with probability $M_1/(2M)$ we pick a
bug and correct it - changing $M_1 \mapsto M_1-1$, and with
probability $c\rho_1 M_1/(2M)$ we pick a ``child'' of a bug which
becomes changed (remember that a bug has on average $c$ children
messages, but only a fraction $\rho_1$ of these becomes changed when
updated under WP) - changing $M_1 \mapsto M_1+1$.

After $2M$ random updates, each message is, on average, visited once.
To obtain time-scales comparable to the parallel update, we therefore
have to rescale time as $t=T/(2M)$, identifying a single random update
with the asymptotically infinitesimal time step $dt = 1/(2M)$. In this
limit, Eqs.~(\ref{eq:rate_update}) can be rewritten as a system of
ordinary differential equations,
\begin{equation}
\frac d{dt} p_t (M_1) = -M_1 p_t(M_1) + (M_1+1) p_t(M_1+1) 
- c\rho_1 M_1 p_t(M_1) + c\rho_1 (M_1-1) p_t(M_1-1)\ .
\end{equation}
For the time evolution of the average number $\overline{M_1(t)} =
\sum_{M_1} M_1 p_t(M_1)$ of bugs we thus find
\begin{eqnarray}
\frac d{dt}\overline{M_1(t)} &=& -\overline{M_1^2}+\overline{M_1(M_1-1)}
-c\rho_1 \overline{M_1^2} + c\rho_1 \overline{M_1(M_1+1)}
\nonumber\\
&=& -(1-c\rho_1) \overline{M_1(t)}\ .
\end{eqnarray}
It decays exponentially with
\begin{equation}
\tau_{rand} = \frac 1 {1-c\rho_{1}}
\end{equation}
and shows thus the same critical behavior as the parallel update.
The main difference appears for $c\to 0$: Whereas the parallel
relaxation time goes to zero, $\tau_{rand}$ approaches one. This
reflects the persistence time that a message is not updated at all:
The fraction of variables which are not selected in $N$ single-spin
updates is $e^{-1}$.

Note that the algorithm as presented in Sec.~\ref{sec:wp_alg} uses a
third update scheme, namely sequential update, which is asynchronous
but sees every message exactly once in $2M$ steps. The analytical
description is more involved than the one of a simple parallel or
random update, but the critical behavior is expected to remain
unchanged. For small $c$, the behavior is further on expected to be
more similar to the one of the parallel update scheme: Since every
message is seen exactly once in $2M$ steps, there are no persistence
effects.

\subsection{Numerical tests}
\label{sec:wp_numerics}

We have performed numerical tests of WP on randomly generated
instances of random graphs at various connectivities $c<e$, for graph
sizes up to $N=10^5$.

To verify the results, we have also applied the {\it leaf-removal
algorithm} which was shown \cite{BauerM2001} to output exact results
exactly in the same connectivity region, and which is the basis of the
proof of correctness of the replica-symmetric result. The algorithm
works as follows: In every step, a leaf (vertex of degree one) is
selected, its neighbor is covered and both vertices are removed from
the graph, as well as all their incident edges. If this algorithm is
able to cover the full graph, the generated VC is a minimum one, but
the algorithm fails if, possibly after some decimation steps, a
leaf-free subgraph emerges.

We have found that both algorithms produce in almost all cases
identical results, i.e. the WP output is thereby shown to be
exact. Also the initial estimate of the VC size after the first
converges of WP, before starting graph decimation, was found to
coincide in the most majority of all cases with the final output. As
discussed above, this is a signal that already the first convergence
of WP leads to exact messages.
\begin{figure}[htb]
  \vspace{1.3cm}
  \begin{center}
    \scalebox{0.5}{\includegraphics{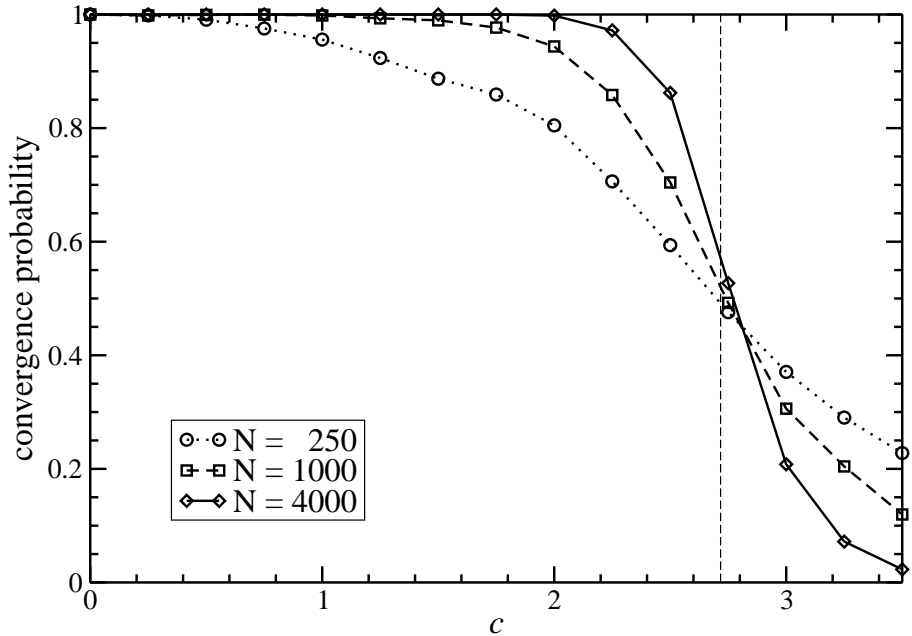}}
  \end{center}
  \caption{Convergence probability of WP as a function of the average
  degree, for graph of $N=250,\ 1000$ resp. 4000 vertices. The symbols
  signify the fraction of graphs, where after 1000 sequential updates
  at least 99\% of the warnings are converged, measured for 10000,
  3000 resp. 1000 sample graphs. The dashed vertical line is situated
  at $c=e$, where WP theoretically ceases to converge.}
  \label{fig:wp_conv}
\end{figure}

The problem of WP is, as discussed before, the slowing down and final
non-convergence if we approach (or exceed) an average degree $c=e$. In
Fig.~\ref{fig:wp_conv}, we have quantified this phenomenon. We have
measured the fraction of graphs of given average degree $c$ (and given
$N$) which, within 1000 sequential WP updates of all $2M$ messages,
are converged on more than 99\% of all messages. In the figure, we see
a clear drop of this probability from almost one to zero in a region
concentrated close to $c=e$. This drop sharpens considerably with
growing graph size $N$, and suggests thus the existence of a sharp
transition in the WP behavior in the thermodynamic limit
$N\to\infty$. Note that, in Fig.~\ref{fig:wp_conv}, this transition
seems to be at a graph degree being slightly larger than $c=e$. This
is a result of the measured quantity: The transition should be found
exactly in $c=e$ when for an arbitrarily large, but finite number of
updates almost all messages are converged -- instead of the test
values used in the generation of Fig.~\ref{fig:wp_conv}.

\section{Survey propagation (SP)}
\label{sec:sp}

We have already mentioned the possibility that the equations of
warning propagation possess a high number of solutions, and none can
be found using a local iterative update scheme. The messages would try
to converge to different, conflicting solutions in different regions
of the graph, and global convergence cannot be achieved. In physics'
language, these different solutions correspond to different
thermodynamic states -- to be understood as clusters of minimal
VCs. Inside such a cluster, any two VCs are connected by at least one
path via other (almost) minimal VCs, which differ stepwise only by a small
number of elements (the number of these different elements stays
finite in the thermodynamic limit). For two minimal VCs selected from
two different clusters, no such connecting paths exist, at least once
an extensive step has to be performed. Note that this distinction is,
from a mathematical point of view, not well-defined for finite graphs
- which are the objects of our algorithms. There can be, however, a
clear separation of distance scales which practically allows for an
identification of solution clusters.
\begin{figure}[htb]
  \begin{center}
    \scalebox{0.5}{\includegraphics{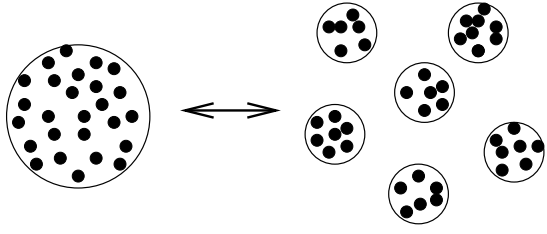}}
  \end{center}
  \caption{Schematic graphical representation of the organization of
    optimal solutions for warning propagation (left side) and for
    survey propagation (right side). For the first case, all solutions
    are collected in one large unstructured cluster (or in a very
    small number of these clusters, as in the case of a ferromagnet),
    corresponding to unbroken replica symmetry. In the second case,
    the set of solutions is clustered into a large number of
    extensively separated subsets. Survey propagation corresponds to
    one step of replica symmetry breaking, where there is no further
    organization of the clusters in larger clusters.}
  \label{fig:wp_vs_sp}
\end{figure}

As already said, warning propagation works well only if there is a
single cluster (or a very small number of clusters) -- corresponding
to the replica symmetric solution. A breaking of the replica symmetry
implies the emergence of clustering in the solution space. This effect
is taken into account by the {\it survey propagation} (SP) algorithm,
as first proposed in \cite{MezardM2002,MezardM2002b}. This algorithm
is equivalent to the first step of replica symmetry breaking, where
the solution clusters show no further organization. If there are
clusters of clusters etc., one has to go beyond survey propagation.

\subsection{The algorithm}

Let us, however, assume a clustering only on one level. Instead of
defining probabilities like $\pi_i$ over the full solution space, we
consider for a moment only one cluster. Inside such a cluster of
minimum VCs, a vertex $i$ may either be always covered (state 1),
never covered (state 0) or sometimes covered and sometimes not (joker
state $*$). This means that, for single clusters, we treat the problem
on the same level as WP.

However, the assignment of this three-valued vertex state may vary
from cluster to cluster.  We now denote by $\hat\pi_i^{(1)}$ the
fraction of clusters where vertex $i$ takes state one, by
$\hat\pi_i^{(0)}$ the fraction of clusters where vertex $i$ takes
state zero, and by $\hat\pi_i^{(*)}$ the fraction of clusters where
vertex $i$ takes the joker state $*$. Analogously we define the cavity
quantities $\hat\pi_{j|i}^{(1)}$, $\hat\pi_{j|i}^{(0)}$ and
$\hat\pi_{j|i}^{(*)}$ on the cavity graph $G_i$. A crucial assumption
of SP is that clusters do not change dramatically by eliminating one
vertex from the graph, i.\,e., by going back and forth between the
full graph and the cavity graphs for different cavities.

Again, we can distinguish the three cases in Fig.~\ref{fig:messages}
of how the variable states propagate inside each solution cluster. A
vertex $i$ of state 0 has to have all neighbors in states $1$ or $*$
on the cavity graph $G_i$; a vertex $i$ of state $*$ has to have
exactly one neighbor of state 0 on the cavity graph; a vertex $i$ of
state 1 has at least two neighbors which have state 0 on the cavity
graph. The statistics over all clusters can now be performed in a very
simple way. The fraction of clusters having vertex $i$ in state 0
which by definition is $\hat\pi_i^{(0)}$ equals the fraction of
solution clusters of the cavity graph $G_i$ where all neighbors are in
a state different from 0, and so on, for the other two states.  This
procedure guarantees the minimization \emph{inside} each
cluster. Note, however, that in clusters belonging to the first case
no vertex has to be added to the minimal VC by stepping from the
cavity graph to the full graph, whereas the VC size increases by one
in the second and third case. The VCs of different clusters thus grow
differently. To optimize \emph{between} clusters, we therefore
introduce a penalty $e^{-y}$ to the last two cases. The resulting
equations are
\begin{eqnarray}
  \label{eq:sp}
  \hat\pi_i^{(0)} &=& C_i^{-1} \prod_{j\in N(i)} (1-\hat\pi_{j|i}^{(0)})
  \nonumber\\
  \hat\pi_i^{(*)} &=& C_i^{-1} e^{-y} \sum_{j\in N(i)} \hat\pi_{j|i}^{(0)}
  \prod_{j'\in N(i)\setminus j} (1-\hat\pi_{j'|i}^{(0)})
  \\
  \hat\pi_i^{(1)} &=& C_i^{-1} e^{-y} 
  \left[1- \prod_{j\in N(i)} (1-\hat\pi_{j|i}^{(0)})
  - \sum_{j\in N(i)} \hat\pi_{j|i}^{(0)}
  \prod_{j'\in N(i)\setminus j} (1-\hat\pi_{j'|i}^{(0)})\right] \ ,
  \nonumber
\end{eqnarray}
and the normalization constant is given by
\begin{equation}
  \label{eq:Ciexpression}
  C_i = e^{-y} \left[ 1- (1-e^y)\prod_{j\in N(i)} (1-\hat\pi_{j|i}^{(0)})
  \right] \ .
\end{equation}
Note that we have  again made an assumption of statistical independence
of the vertices $j$ on the cavity graph. This assumption enters on two
levels: First inside the cluster, when we say that $j$ vertices of
state $*$ can be covered simultaneously in a minimum VC of the cavity
graph; and second in between clusters, when we factorize the joint
probabilities in the upper expression.

Analogous equations are valid for the iteration of the
cavity quantities, where again the influence of the cavity site has to
be taken out:
\begin{eqnarray}
  \label{eq:spB}
  \hat\pi_{i|l}^{(0)} &=& C_{i|l}^{-1} \prod_{j\in N(i)\setminus l} 
  (1-\hat\pi_{j|i}^{(0)})
  \nonumber\\
  \hat\pi_{i|l}^{(*)} &=& C_{i|l}^{-1} e^{-y} \sum_{j\in N(i)\setminus l} 
  \hat\pi_{j|i}^{(0)}
  \prod_{j'\in N(i)\setminus \{j,l\}} (1-\hat\pi_{j'|i}^{(0)})
  \nonumber\\
  \hat\pi_{i|l}^{(1)} &=& C_{i|l}^{-1} e^{-y} 
  \left[1- \prod_{j\in N(i)\setminus l} (1-\hat\pi_{j|i}^{(0)})
  - \sum_{j\in N(i)\setminus l} \hat\pi_{j|i}^{(0)}
  \prod_{j'\in N(i)\setminus \{j,l\}} (1-\hat\pi_{j'|i}^{(0)})\right] 
  \nonumber\\
  C_{i|l} &=& e^{-y} \left[ 1- (1-e^y)\prod_{j\in N(i)\setminus l} 
    (1-\hat\pi_{j|i}^{(0)})
  \right] \ .
\end{eqnarray}
These are the equations for \emph{survey propagation}.  To solve them,
one has to first initialize the cavity quantities arbitrarily and
update them iteratively according to the second set of equations. Once
convergence is reached, the $\hat\pi_i^{(\cdot)}$ can be simply
evaluated from the first set of equations. Note also that the SP
equation for the cavity quantities close in the $\hat\pi_{j|i}^{(0)}$
alone:
\begin{equation}
\hat\pi_{i|l}^{(0)} = \frac
{\prod_{j\in N(i)\setminus l} (1-\hat\pi_{j|i}^{(0)})}
{e^{-y} \left[ 1- (1-e^y)\prod_{j\in N(i)\setminus l} (1-\hat\pi_{j|i}^{(0)}) 
\right]} \ .
\label{eq:pi_iter_sp}
\end{equation}

A note on the selection of the re-weighting parameter $y$ is necessary:
Finite values of $y$ focus on {\it local minima} of the complex
landscape $X(U)=|U|$ defined over all VCs, i.e.~to VCs of cardinality
which cannot be decreased by changing only a finite part of $U$. One
would thus expect naively that minimal VCs are obtained in the limit
$y\to\infty$. As we will see in the next section, the SP solution
carries, however, sensible physical information only in a limited
interval $y\in [0,y^*]$. It is therefore necessary to work directly
with finite $y$-values.

The knowledge of all $\hat\pi_i^{(\cdot)}$ does not allow us to
directly create a (locally) minimal vertex cover. It is impossible to
deduce a joint probability distribution of all vertices from the
knowledge of the marginal single-vertex probabilities
only. Nevertheless, some useful knowledge can be drawn directly from
these quantities. In particular we may estimate the VC size by
\begin{equation}
X(y) = \sum_{i\in V} \left( \hat\pi_i^{(1)} + \frac 12
\hat\pi_i^{(*)} \right) \ .
\label{eq:X_y}
\end{equation}
As in the replica-symmetric case of WP, we have assumed that vertices
carrying value $*$ are, on average, half covered and half
uncovered. At this point, this is a pure conjecture which, however,
will be justified in the next section [see Eq.~(\ref{eq:energyconsist})].

To actually construct a minimum vertex cover (or an approximation due
to the non-exactness of SP because of, e.g., a finite value of $y$,
cycles in the graph or more levels of cluster organization), we have
to resort again to an iterative decimation scheme.  In every step, the
$\hat\pi_i^{(\cdot)}$ are calculated, one vertex of large
$\hat\pi_i^{(1)}$ is selected and covered. It is removed from the
graph together with all incident edges, and the $\hat\pi_i^{(\cdot)}$
are reiterated on the decimated graph. This procedure is iterated
until no uncovered edges are left, i.\,e., until a proper VC is
constructed. Slightly different schemes of selection heuristics can be
applied (select a vertex of high $\hat\pi_i^{(0)}$, uncover it, cover
all neighbors, and decimate the graph, or take into account also the
value of $\hat\pi_i^{(*)}$). All these heuristic rules are equally
valid in the sense that, if SP is exact on a graph, they all produce
one minimum VC of the same size. For real graphs, however, where the
results of the SP equations are to be considered as approximations of
the actual quantities defined over the set of solutions, different
heuristic choices may result in VCs of different sizes. Within the
numerical experiments described below, we have, however, found no
preferable selection heuristic, and the fluctuations from one
heuristic to another were small compared with the VC size.

\subsection{The complexity of clusters}
\label{sec:complexity}

Different values of the re-weighting parameter $y$ lead to a
concentration of the partition sum (or, equivalently, the solution of
the SP equations) to clusters of vertex covers of different (locally
minimal) size. The {\it complexity} $\Sigma(X)$, or {\it
configurational entropy}, measures the logarithm of the number ${\cal
N}_{cl}(X)$ of clusters of given VC size $X$. We introduce the
generalized thermodynamic potential $\Phi(y)$ as the Legendre
transform of the complexity,
\begin{equation}
e^{-y \Phi(y)} = \sum_{X=0}^N \exp\left\{ - y X + \Sigma(X) \right\}\ .
\label{eq:Phi}
\end{equation}
According to the general procedure of the cavity method in diluted
systems \cite{MezardM2001}, this potential can be decomposed into site and
link contributions,
\begin{equation}
\Phi(y) = \sum_{i\in V} \Delta\Phi_i(y) - \sum_{\{i,j\}\in E}
\Delta\Phi_{i,j}(y)\ .
\end{equation}
These contributions can be determined by adding a vertex / an edge to
the graph:
\begin{eqnarray}
e^{-y \Delta\Phi_i(y)} &=& \prod_{j\in N(i)} (1-\hat\pi_{j|i}^{(0)}) +
\left(1-\prod_{j\in N(i)} (1-\hat\pi_{j|i}^{(0)}) \right) e^{-y}
\nonumber\\ &=& e^{-y} + (1-e^{-y}) \prod_{j\in N(i)}
(1-\hat\pi_{j|i}^{(0)}) \nonumber\\ e^{-y \Delta\Phi_{i,j}(y)} &=&
1-(1-e^{-y})\hat\pi_{j|i}^{(0)}\hat\pi_{i|j}^{(0)}
\end{eqnarray}
where, as in the derivation of the SP equations, one has to take care
separately of the cases where the VC size remains unchanged under
vertex or link addition, or increases by one. Having solved the SP
equations for the $\hat\pi_{i|j}^{(0)}$, the potential becomes easy to
calculate,
\begin{equation}
-y \Phi(y) = \sum_{i\in V} \ln \left(e^{-y}   
+ (1-e^{-y}) \prod_{j\in N(i)}(1-\hat\pi_{j|i}^{(0)}) \right) 
 - \sum_{\{i,j\}\in E} \ln\left( 1-(1-e^{-y})\hat\pi_{j|i}^{(0)}
\hat\pi_{i|j}^{(0)} \right) \ .
\label{eq:Phi2}
\end{equation}
Approximating the sum in Eq.~(\ref{eq:Phi}) by the saddle point method
(valid for $N\gg 1$), we see that the complexity can be calculated via
\begin{equation}
\Sigma(y) = \Sigma(X(y)) = y ( X(y)-\Phi(y) )\ ,
\label{eq:Phi3}
\end{equation}
where $X(y)$ is given in Eq.~(\ref{eq:X_y}) in dependence of the SP
solution. The function $X(y)$ can also be determined directly from the
potential $\Phi(y)$ via $X(y) = \Phi(y)+y \Phi'(y)$. The numerical
observation that both expressions for $X(y)$ coincide is a strong
justification for the ratio 2 used in Eq.~(\ref{eq:X_y}) between the
number of all unfrozen vertices and the number of simultaneously
covered unfrozen vertices.

The complexity $\Sigma$ is defined as the logarithm of the cluster
number, i.e.~in the presence of at least one cluster it takes
necessarily a non-negative value. This defines a range $y\in(0,y^*)$
where the SP equation provide a potentially sensible solution, with
$y^*$ given by the marginality condition $\Sigma(y^*)=0$. For higher
$y$, the predicted complexities become negative -- corresponding thus
to un-physical solutions of the SP equations. We see that the naive
expectation that $y\to\infty$ leads to minimal VCs is thus
inconsistent, the best possible estimate for the minimal VC size we
can obtain at the level of SP (one-step replica symmetry breaking) is
thus given by $X(y^*)$ \cite{ZhouH2003}. Note that this observation,
in replica theory, corresponds to the usual optimization of the
replicated free energy over the replica symmetry breaking parameter
\cite{MePaVi,Mo}. Note also that the existence of a finite $y^*$ is a
clear signal for the existence of more than one step of replica
symmetry breaking, and the SP results can only be expected to be
approximations to true minimal VCs.

\subsection{Stability of the fixed point under SP iteration}
\label{sec:sp_stab}

It is, however, not clear if SP converges at all in the
replica-symmetry broken phase. To investigate this question, we
consider the behavior of the solution of Eq.~(\ref{eq:pi_iter_sp}) under
small perturbations. Note that the situation here is different from
the bug proliferation picture used in order to analyze the stability of
WP fixed points: The messages now are real numbers and thus small
perturbations are possible even on the level of a single message.

Let us therefore imagine that we start a set of experiments, with
initial conditions $\pi^{(0)}_{i|l}$ distributed around the SP
solution $\hat \pi^{(0)}_{i|l}$ according to some narrow distribution
\begin{equation}
f_{i|l} ( \pi^{(0)}_{i|l} ) = \frac 1{\sqrt{2 \pi}
\varepsilon_{i|l} } \exp\left\{ - \frac
{\left[ \pi^{(0)}_{i|l}-\hat \pi^{(0)}_{i|l}\right]^2}
{2\varepsilon_{i|l}^2}\right\}
\end{equation}
of link-dependent widths $\varepsilon_{i|l}\ll 1$. After one iteration
of SP, the messages are distributed according to (for simplicity we
have skipped the superscript $(0)$)
\begin{equation}
f'_{i|l} ( \pi_{i|l} ) = \int \prod_{j\in N(i)\setminus l}
\left[ d \pi_{j|i}  f_{j|i} (\pi_{j|i}) \right]
\delta\left(\pi_{i|l} -\tilde \pi_{i|l}( \{\pi_{j|i}\} )
\right)
\end{equation}
with the update rule $\tilde \pi$ given by  Eq.~(\ref{eq:pi_iter_sp}).
We expand this update function around the SP fixed point,
\begin{equation}
\tilde \pi_{i|l}( \{\pi_{j|i}\} ) = \hat \pi_{i|l}^{(0)}
+ \sum_{j\in N(i)\setminus l} \frac{\partial \hat\pi_{i|l}^{(0)}}{\partial
\hat\pi_{j|i}^{(0)}} (\pi_{j|i}-\hat\pi_{j|i}^{(0)})
+ \frac 12 \sum_{k,j\in N(i)\setminus l} 
\frac{\partial^2 \hat\pi_{i|l}^{(0)}}{\partial
\hat\pi_{k|i}^{(0)}\partial\hat\pi_{j|i}^{(0)} } (\pi_{k|i}-\hat\pi_{k|i}^{(0)})
(\pi_{j|i}-\hat\pi_{j|i}^{(0)})+{\cal O}(\varepsilon^3)\ .
\end{equation}
The mean of the updated message is given by
\begin{eqnarray}
\langle \pi_{i|l} \rangle' &=& \int \prod_{j\in N(i)\setminus l}
\left[ d \pi_{j|i}  f_{j|i} (\pi_{j|i}) \right] 
\tilde \pi_{i|l}( \{\pi_{j|i}\} )\nonumber\\
&=& \hat \pi_{i|l}^{(0)} + \frac 12 \sum_{j\in N(i)\setminus l} 
\frac{\partial^2 \hat\pi_{i|l}^{(0)}}{(\partial \hat\pi_{j|i}^{(0)})^2} 
\varepsilon_{j|i}^2\ ,
\end{eqnarray}
and its change is negligible with respect to the width of the
distribution.  The second moment, on the other hand, behaves as
\begin{eqnarray}
\langle \pi_{i|l}^2 \rangle' &=& \int \prod_{j\in N(i)\setminus l}
\left[ d \pi_{j|i}  f_{j|i} (\pi_{j|i}) \right] 
\tilde \pi_{i|l}^2( \{\pi_{j|i}\} )\nonumber\\
&=& (\hat \pi_{i|l}^{(0)})^2 + \hat \pi_{i|l}^{(0)} \sum_{j\in N(i)\setminus l} 
\frac{\partial^2 \hat\pi_{i|l}^{(0)}}{(\partial \hat\pi_{j|i}^{(0)})^2} 
\varepsilon_{j|i}^2 + \sum_{j\in N(i)\setminus l} 
\left(\frac{\partial \hat\pi_{i|l}^{(0)}}{\partial \hat\pi_{j|i}^{(0)}}
\right)^2  \varepsilon_{j|i}^2 \ .
\end{eqnarray}
We find thus that the variance of the updated distribution behaves as
\begin{equation}
{\varepsilon'_{i|l}}^2 = \langle \pi_{i|l}^2 \rangle' - 
{\langle \pi_{i|l} \rangle'}^{2} = \sum_{j\in N(i)\setminus l} 
T_{i|l,j|i}\ \varepsilon_{j|i}^2
\label{eq:iter_eps} 
\end{equation}
with
\begin{equation}
T_{i|l,j|i} = \left(\frac{\partial \hat\pi_{i|l}^{(0)}}{\partial
\hat\pi_{j|i}^{(0)}}\right)^2 = \left(\frac { e^{-y} \prod_{k\in N(i)\setminus
\{j,l\}} (1-\hat\pi_{k|i}^{(0)})} { \left[e^{-y} + (1-e^{-y})
\prod_{k\in N(i)\setminus j} (1-\hat\pi_{k|i}^{(0)}) \right]^2} 
\right)^2 \ .
\label{eq:Tmatrix}
\end{equation}
The (in)stability of this equation is related to the largest eigenvalue
$\lambda_{max}$ of the matrix $(T_{i|l,j|i})$, only if $\lambda_{max}$
is smaller than one the perturbations $f_{i|l} ( \pi^{(0)}_{i|l} )$
of the SP solution contract exponentially.

Note that this type of stability of the SP fixed point is known in the
literature under the name ``type-one instability''
\cite{MoRi,MeMeZe} and can be related to the appearance of
more than one step of replica symmetry breaking, more precisely to the
fragmentation of the solution clusters in sub-clusters. It is not the
only type of instability of the one-step replica-symmetry-broken
solution with respect to more steps, an alternative scheme would be
the accumulation of clusters into clusters of clusters (``type-two
instability''). The latter instability leads, however, not to an
iterative instability of the SP equations itself, i.e. the later can
be used even if not being physically exact. This is what happens in
the case of VC \cite{ZhouH2003}.

\subsection{Numerical tests}
\label{sec:sp_num}

\subsubsection{The size of the constructed VC}
\label{sec:sp_num_size}

\begin{figure}[htb]
  \vspace{1cm}
  \begin{center}
    \scalebox{0.4}{\includegraphics{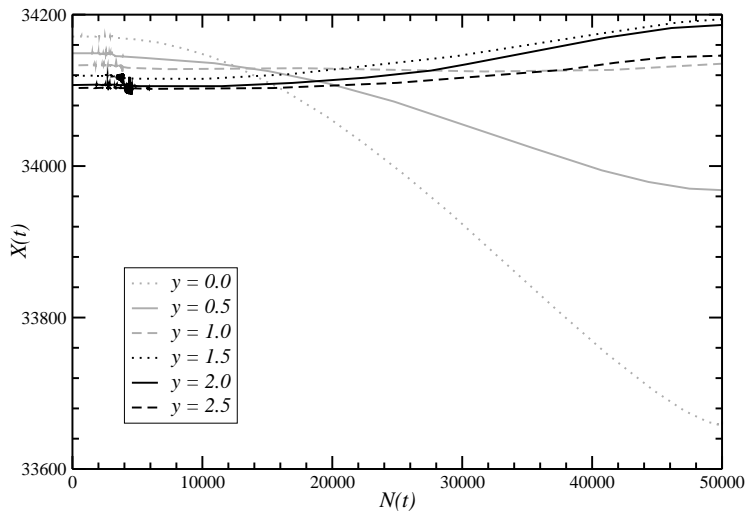}}
  \end{center}
  \caption{Trajectories of graph decimation for a single graph with
  $N=50\ 000$ and $c=10$. We plot the VC size as estimated by SP, as a
  function of the vertex number in the remaining graph. The decimation
  proceeds from the right to the left, i.e.~from the initial $N=50\
  000$ toward zero. The fact that the estimate changes under
  application of the reduction process results from the approximate
  nature of the SP messages.}
  \label{fig:traj}
\end{figure}

In order to check the performance of SP, we have tested it on single
samples of random graphs of medium to large size. In
Fig.~\ref{fig:traj}, we concentrate on a single graph of $N=50\ 000$
vertices and average degree $c=10$. The data reported in
Fig.~\ref{fig:traj} are quantitatively comparable to other graphs with
the same parameters, and qualitatively with graphs of other sizes and
connectivities. We show trajectories of the estimated VC size during
graph reduction, as a function of the number of vertices which are
still in the reduced graph, for various values of the re-weighting
parameter $y$. We see that the initial variability of the estimates is
much larger than the difference in output. Even the worst performing
case, $y=0$, outputs a VC of 34 171 vertices compared to the minimal
found one with $X=34\ 104$. This similarity is due to the fact that
the ranking of the vertices with respect to the SP results depends
only weakly on $y$, whereas the messages itself change considerably -
and thus the corresponding predictions of the VC size. Close to the
end of the curves, there are some striking fluctuations in the VC
size. In this region, SP was not able to converge to a fixed point,
and the non-converged solution was used. This non-convergence of SP
may be related to the critical slowing down of SP at the phase
boundary when the solution space of the minimal VC problem transits
between the two schemes of Fig.~\ref{fig:wp_vs_sp} (see
Sec.~\ref{subsec:relaxation}).  After an interval of these
fluctuations, the SP solution collapses to the WP solution, i.e.~we go
from the replica-symmetry broken, clustered phase to the
replica-symmetric, unclustered phase. Note that the performance of SP
improves with increasing values of $y$, as long as it converges in the
most majority of the decimation steps. The region of non-convergence,
however, grows with $y$.

To circumvent this problem, we have introduced a version of SP with
adaptive $y$-values. We start SP with a relatively large $y$, and
whenever the convergence time exceeds a certain threshold (we have
used, e.g., 100 sequential updates of all messages), the value of $y$
is decreased (we have, e.g., multiplied it by 0.9). As a result, the
trajectory of predicted VC sizes is smoothened, and the algorithm
automatically tends toward the lowest found VC sizes.

As already mentioned, the original estimate for $X$ varies a lot with
$y$, it is even non-monotonous. Whereas the value for $y=0$ is
substantially smaller than the smallest constructed VCs, there is a
local maximum which is larger than the constructed VCs. It is,
however, astonishing that the extrapolated value at $y^*$, where the
complexity $\Sigma$ vanishes, is extremely close to the finally
constructed value: $34\ 090\pm10$ compared to $34\ 104$. This is even
more astonishing since we do not reach convergence of SP at $y^*$ for
$c=10$, see the discussion in the next sub-section.

\begin{figure}[htb]
  \vspace{1cm}
  \begin{center}
    \scalebox{0.4}{\includegraphics{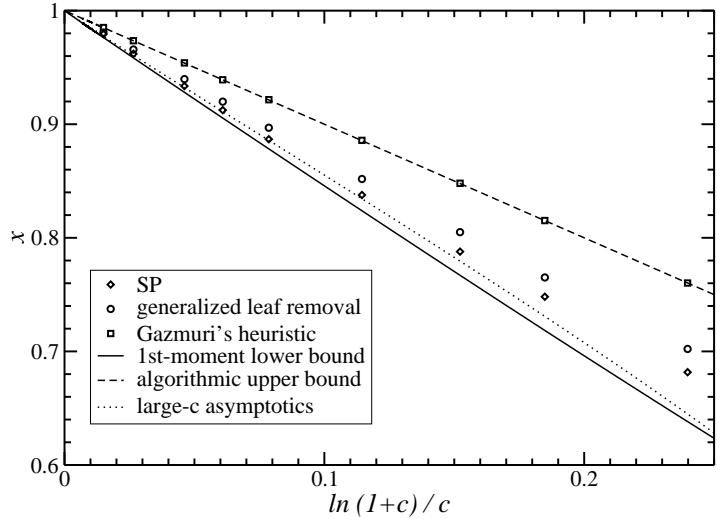}}
  \end{center}
  \caption{Numerical results of SP run on graphs of high, but finite
  average degree $c\in[10,400]$. As a comparison, we have added the
  results of two local algorithms (Gazmuri's heuristic and GLR), and
  rigorous bounds on the asymptotic average behavior for random graphs
  as well as the exact large-$c$ asymptotics. SP clearly outperforms
  local algorithms, and is close to the asymptotically exact value.}
  \label{fig:vc_large_c}
\end{figure}

To see the behavior of SP in the full range of average degrees, we
have systematically scanned the $c$-interval $[10,400]$, as can be
seen in Fig.~\ref{fig:vc_large_c}. The graph size for these high
connectivities range up to $N=6400$ (note that in this case up to
$cN=2\ 560\ 000$ messages have to be handled). The results for fixed
$c$ and various $N$ were extrapolated to their asymptotic value at
$N\to\infty$, in order to be comparable to analytical results and to
the performance of local algorithms. We see that SP performs much
better than the local heuristics, and its behavior is consistent to
the exact large-$c$ behavior found by Frieze \cite{Fr}. For the
comparison we have used two local algorithms: The first one is a
simple heuristic by Gazmuri \cite{Gaz}, where in every step a vertex
is selected randomly, all its neighbors are covered, and the covered
edges are removed from the graph. The second heuristic is a
generalization of leaf removal \cite{Wei} working also beyond average
degree $c=e$, but not guaranteeing minimality of the constructed VC any
more. The algorithm selects in every step a vertex of minimal degree,
covers its neighbors and removes all considered vertices and covered
edges. If the algorithm never needs to select vertices of degree
exceeding one, it reduces to leaf removal. As already said, SP
outperforms the local algorithms.

A drawback for all $c$-values is, however, that the algorithm does not
work at high values of $y$, which, seen the derivation of the SP
equations, should bring us closest to a minimal VC.

\subsubsection{On the iterative stability of SP}
\label{sec:sp_num_stab}

Running SP for different values of the re-weighting parameter $y$, we
observe that it converges very fast for small $y$ (and $c\neq e$), but
it does not converge at all for large $y$. As a first impression, it
seems therefore useless to check the stability of the SP solution via
the eigenvalues of the stability matrix $(T_{i|l,j|i})$.  The solution
itself is found via iteration of Eq.~(\ref{eq:pi_iter_sp}) starting
from a random initial condition, {\it i.e.}~if this iteration
converges, the solution is automatically stable. On the other hand, it
is much harder to extrapolate precisely the point where the
convergence time diverges - instead of identifying this point by
$\lambda_{max}\to 1$.

Technically the eigenvalue can be determined in a way inspired by the
message-passing procedure itself: We randomly initialize all
$\varepsilon_{i|j}$ to non-negative values, and update it according to
Eq.~(\ref{eq:iter_eps}). Then, we renormalize the vector dividing it
by $\sqrt{\sum_{(i,j)\in
E}({\varepsilon_{i|j}}^2+{\varepsilon_{j|i}}^2)}$.  This is repeated
until convergence of the procedure is reached, and $\lambda_{max}$
equals the asymptotic renormalization factor, see \cite{SeMa} for an
analogous approach to testing the stability of a replica symmetric
solution in the problem of counting graph loops.

\begin{figure}[htb]
  \vspace{1cm}
  \begin{center}
    \scalebox{0.4}{\includegraphics{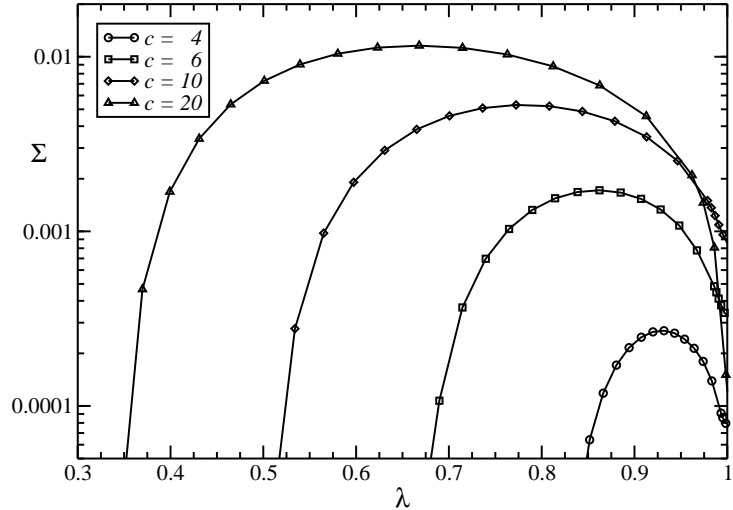}}
  \end{center}
  \caption{Stability of the SP solution: The complexity $\Sigma$ is
  plotted versus the largest eigenvector $\lambda_{max}$ of the
  stability matrix $(T_{i|l,j|i})$ defined in Eq.~(\ref{eq:Tmatrix}),
  for various values of the average graph degree $c$. All data are
  produced from graphs with $N=10\ 000$ vertices, averaged over 10
  samples. Error bars both in $\lambda$ and $\sigma$ are smaller than
  the symbol size.}
  \label{fig:stability_sp}
\end{figure}

The results of the numerical tests are shown in
Fig.~{\ref{fig:stability_sp}}. For the $c$-values displayed there, we
find that the SP solution is stable against small perturbations in the
vicinity of $y=0$, but $\lambda_{max}$ starts to grow right away. For
all the displayed values, we also find that $\lambda_{max}$ approaches
one at positive complexity, {\it i.e.}~at values $y<y^*$. At the
$y$-value which, in the ensemble average, the 1RSB result is expected
to be most precise compared to the exact value, SP does not even
converge on the single sample.

This changes for $c>20.4$. At this point, the instability threshold
coincides precisely with the zero-complexity point corresponding to
minimal VCs. At even higher value, SP thus converges at
$y=y^*$. However, this does not necessarily mean that we can do all
the decimation process efficiently at the initial $y^*$, after
decimation of a $c$-dependent fraction of the graph SP starts to
diverge even for large $c$.

\section{From survey propagation to typical properties on random graphs}
\label{sec:mf}

In Sec.~\ref{sec:wp_to_rs} we have seen, that it is possible to
average the solution of warning propagation over random graphs of
average vertex degree $c$, and to recover the replica-symmetric
results of \cite{WeigtM2000} in the thermodynamic limit. In analogy,
the equations of survey propagation can be used to reproduce and
extend the results of \cite{ZhouH2003} on the typical properties of
minimal VCs under the assumption of one step of replica-symmetry
breaking (1RSB), i.e., to translate the probabilistic-algorithmic
approach on single graph instances to a statistical-physics approach
with the graph randomness playing the role of the quenched disorder.

As already explained above, the computational hardness of the VC
problem results from the fact that the landscape of VC sizes over the
space of all vertex covers may become rough. It may contain, in
particular, many local minima: A VC $U$ is considered to be locally
optimal if all VCs differing only in a finite number of vertices are as
least as large as $U$. Such local minima are expected to act as traps
for many local search algorithms, which therefore are unable to find
globally optimal VCs. If 1RSB is considered, not only the minimal VCs are
assumed to be clustered (cf. Fig.~\ref{fig:wp_vs_sp}), but also an
exponential number of clusters of locally optimal VCs are expected to
appear. The total number of such VC clusters of cardinality $X=xN$ in
a given graph $G$ is denoted as $\Omega_{G}(X)$. The complexity of
graph $G$ at VC density (or $`$energy density') $x$ is defined as
\begin{equation}        
 \Sigma_G(X)=  \frac 1N \ln \Omega_{G}(X) \ ,       
 \label{eq:complexity}  
\end{equation}
with respect to Sec.~\ref{sec:complexity} we have renormalized the
complexity by a factor $1/N$ to assure a sensible thermodynamic
limit. The complexity $\Sigma_{G}(X)$ is expected to be self-averaging
\cite{MePaVi}: When $N$ is increased, the complexity of randomly drawn
graphs $G$ approaches asymptotically the mean value averaged over the
whole graph ensemble. Technically, the interesting quantity is thus
\begin{equation}
 \Sigma(c,x)= \lim_{N\to\infty} \frac 1N \langle \ln \Omega_{G}(xN)
 \rangle_{G} \ ,
 \label{eq:mean_entropy}
\end{equation}
where $\langle \cdots \rangle_{G}$ denotes average over all random
graphs $G$ of fixed parameters $N$ and $c$, since this value is found
almost surely also in very large random graphs.

The partial derivative of $\Sigma(c,x)$ with respect to $x$ is denoted
as
\begin{equation}
 \label{eq:y}
  y={\partial \Sigma(c,x) \over \partial x} \ .
\end{equation}
The above equation gives an implicit relationship between $y$ and the
VC density $x$. We can define a generalized free energy density at
given value of $y$ and $c$ via the Legendre transform
\begin{equation}
 \label{eq:f}
  \phi(c, y)= x(c, y) - {\Sigma\bigl( c, x(c, y)\bigr) \over y} \ ,
\end{equation}
in complete analogy to the single-graph quantity $\Phi$ in
Eq.~(\ref{eq:Phi}).

The parameter $y$ formally corresponds to an inverse temperature
$\beta$ in an ordinary statistical physics system, with the difference
that microscopic configurations are replaced by clusters of locally
optimal VCs. It can be used  to control the mean VC density $x$ of
our artificial statistical-physics system. This is in fact done in the
SP equations, as we will see below, $y$ is exactly the re-weighting
parameter introduced before.

All this holds true as long as the relative VC size $x = X/N$ is such
that $\Omega_G(xN) \gg 1$ for a typical random graph $G$, i.e.,
$\Sigma(c,x)\geq 0$. When $\Sigma(c,x)<0$, a typical random graph $G$
has no optimal VC clusters of density $x$. The largest allowable value
$y = y^*$ is thus located at the point where $\Sigma\bigl( c,x(c, y^*)
\bigr)=0$. This point also corresponds to the best 1RSB estimate of
the globally minimal VC size $x(c,y^*)$ of a typical random graph,
cf. the discussion at the end of Sec.~\ref{sec:complexity}.

\subsection{The cavity equation}

As already discussed, for each cluster of VCs, vertices can be
decorated by a three-state variable: It assumes the value 1, if the
vertex belongs to all VCs in the cluster, it takes the value * if it
belongs to some but not all VCs, and the value 0 if it belongs to no
VCs of the cluster. Let us also recall the notation $\hat\pi_i^{(0)}$,
$\hat\pi_i^{(*)}$, and $\hat\pi_i^{(1)}$
$(=1-\hat\pi_i^{(0)}-\hat\pi_i^{(*)})$ for the probability that vertex
$i$ takes the corresponding value in a randomly selected locally
optimal VC at given $y$. The values
$\vec\pi_i=(\hat\pi_i^{(0)},\hat\pi_i^{(*)},\hat\pi_i^{(1)})$
fluctuate from vertex to vertex, and the main aim of the cavity method
is to describe their distribution in a self-consistent way.

Suppose one already knows $\vec{\pi}_i$ for each vertex $i$ of a
random graph $G$ with $N$ vertices. Now add a new vertex (say vertex
$0$) and connect it to $k$ randomly chosen vertices (say
$j=1,2,\ldots, k$) of graph $G$. The integer $k$ is determined
according to the Poisson distribution $f_c(k)=e^{-c} c^k / k!$.  After
vertex $0$ and the $k$ edges are added, a new graph $G^\prime$ of
$N+1$ vertices is constructed.  Under the assumption of statistical
independence of the vertices $j=1,2,\ldots,k$ in graph $G$ [cf.~the
comment below Eq.~(\ref{eq:Ciexpression})], one can write down the
following equations for $\vec\pi_0$:
\begin{eqnarray}
  \hat\pi_0^{(0)}&=& {\prod\limits_{j=1}^{k} (1-\hat\pi_j^{(0)}) 
    \over e^{-y}+(1-e^{-y}) 
    \prod\limits_{j=1}^{k} (1-\hat\pi_j^{(0)})} \ , \label{eq:pi00} \\
  \hat\pi_0^{(*)}&=& {e^{-y} \sum\limits_{j=1}^{k} \hat\pi_j^{(0)} 
    \prod\limits_{l\neq j}  (1-\hat\pi_l^{(0)}) \over e^{-y}+(1-e^{-y}) 
    \prod\limits_{j=1}^{k} (1-\hat\pi_j^{(0)}) } \ , \label{eq:pi0*} \\
  \hat\pi_0^{(1)} &=& 1-\hat\pi_0^{(0)}-\hat\pi_0^{(*)} \ .
  \label{eq:pi01}
\end{eqnarray}
Assuming furthermore that the statistical properties of the $\vec\pi$
do not change drastically by adding the new vertex,
Eq.~(\ref{eq:pi00}) allows us to write down the following
self-consistent equation governing the probability distribution of
$\hat\pi_0^{(0)}$:
\begin{equation}
  P(\hat\pi_0^{(0)}) = f_c(0) \delta_{\hat\pi_0^{(0)},1}+
  \sum\limits_{k=1}^{\infty}f_c(k) \prod\limits_{l=1}^{k} \left[\int
  {\rm d} \hat\pi_l^{(0)} P(\hat\pi_l^{(0)})\right] \delta\left(
  \hat\pi_0^{(0)}-{\prod\limits_{l=1}^{k} (1-\hat\pi_l^{(0)}) \over e^{-y} +
  (1-e^{-y}) \prod\limits_{l=1}^{k} (1-\hat\pi_l^{(0)})} \right) \ .
  \label{eq:pi00distr}
\end{equation}
This equation can be numerically solved with very high precision using
a standard algorithm of population dynamics. Note also the equivalence
of the update rule in the Delta function to
Eq.~\ref{eq:pi_iter_sp}. One can in fact estimate $P(\hat\pi_0^{(0)})$
also by first generating a huge random graph, iterating SP on it and
than calculating the histogram of all messages.

\subsection{VC density and complexity}

Also the VC density is self-averaging. When the graph size $N$ is
sufficiently large, the VC density of a typical graph $G$ is almost
independent of the microscopic details of $G$; it only depends on the
statistical properties of the graph ensemble represented by the mean
vertex degree $c$. At fixed value of the re-weighting parameter $y$,
also this mean VC density $x(c,y)$ can be calculated using the cavity
method. The graph $G^\prime$ as generated in the preceding subsection
has $N+1$ vertices and mean vertex degree $c^\prime= 2(M+k)/(N+1) = c
+ (2k-c)/(N+1)$. The expectation of the VC density of $G^\prime$ and
that of the graph $G$ are related by
\begin{equation}
  (N+1)\ x(c^\prime, y) = N \ x(c, y) + 1 - \hat\pi_0^{(0)} \ .
 \label{eq:enew01}
\end{equation}
Expanding $x(c,y)$ around $c$, and keeping only the non-vanishing
terms in the thermodynamic limit, we obtain 
\begin{equation}
        \label{eq:enew02}
         x(c,y)+c {\partial x(c, y) \over \partial c}
                 =1- \langle \hat\pi_0^{(0)} \rangle_{G^\prime} \ .
\end{equation}

To obtain an expression for ${\partial x(c,y) / \partial c}$, we add a
new edge between two randomly chosen vertices (say vertex $j$ and $l$)
of the old graph $G$ and thus construct a new graph
$G^{\prime\prime}$. This new graph has mean vertex degree
$c^{\prime\prime}=c+2/N$. Averaged over all the locally optimal VC
clusters at fixed re-weighting parameter $y$, the mean increase in
VC density due to addition of edge $(j, l)$ is
\begin{equation}
        \label{eq:de02} 
        {e^{-y}  \hat\pi_j^{(0)} \hat\pi_l^{(0)} 
         \over 1-(1-e^{-y}) \hat\pi_j^{(0)} \hat\pi_l^{(0)} }  \ ,
\end{equation}
since it results from the case that both end vertices $j$ and $l$ are
uncovered in the corresponding cluster. In other words, we have
\begin{equation}
         \label{eq:enew03}
         N x(c^{\prime\prime},y)=N x(c,y)
         +{e^{-y} \hat\pi_j^{(0)} \hat\pi_l^{(0)} 
                  \over 1-(1-e^{-y}) \hat\pi_j^{(0)} \hat\pi_l^{(0)} } \ ,
\end{equation}
which leads to the expression
\begin{equation}
        \label{eq:enew04}
         {\partial x(c,y) \over \partial c} = { 1 \over 2}
         \left\langle {e^{-y} \hat\pi_j^{(0)} \hat\pi_l^{(0)} 
          \over 1-(1-e^{-y}) \hat\pi_j^{(0)} \hat\pi_l^{(0)} } 
	 \right\rangle_{G^{\prime\prime}} \ .
\end{equation}

Combining Eq.~(\ref{eq:enew02}) and Eq.~(\ref{eq:enew04}) we finally obtain that
\begin{eqnarray}
  x(c,y)&=& 1-\langle \hat\pi_0^{(0)} \rangle_{G^\prime}
  -{c \over 2} \left\langle {e^{-y} \hat\pi_j^{(0)} \hat\pi_l^{(0)} 
      \over 1-(1-e^{-y}) \hat\pi_j^{(0)} \hat\pi_l^{(0)} } \right\rangle_{G^{\prime\prime}}
        \nonumber \\
        &=& 1- \int_{0}^{1} {\rm d} \hat\pi^{(0)} P(\hat\pi^{(0)}) \hat\pi^{(0)}  
        -{c \over 2} \int_{0}^{1} {\rm d} \hat\pi_1^{(0)} P(\hat\pi_1^{(0)})
        \int_{0}^{1} {\rm d} \hat\pi_2^{(0)} P(\hat\pi_2^{(0)})
        {e^{-y} \hat\pi_1^{(0)} \hat\pi_2^{(0)}   
          \over 1-(1-e^{-y}) \hat\pi_1^{(0)} \hat\pi_2^{(0)}} \ ,
        \label{eq:enew06}
\end{eqnarray}
where we have used another time the argument that the change in the
histogram $P(\hat\pi{(0)})$ due to vertex or edge addition is neglectable
in the thermodynamic limit.

The first line of Eq.~(\ref{eq:enew06}) is consistent with the
analogue Eq.~(\ref{eq:X_y}) for a single graph. To see this, we notice
that
\begin{eqnarray}
  {c\over 2} \left\langle {e^{-y} \hat\pi_{j}^{(0)} \hat\pi_{l}^{(0)} \over 
      1-(1-e^{-y})\hat\pi_{j}^{(0)} \hat\pi_{l}^{(0)}} \right\rangle
  &=& {1\over 2} \left\langle \sum_{j\in N(l)} {e^{-y} \hat\pi_{j}^{(0)} 
      \hat\pi_{l}^{(0)} \over 1-(1-e^{-y})\hat\pi_{j}^{(0)} \hat\pi_{l}^{(0)}} 
      \right\rangle \nonumber \\
  &=&{1 \over 2} \left\langle \sum\limits_{j\in N(l)} \frac{ e^{-y} \hat\pi_{j}^{(0)}
      {\prod\limits_{i\in N(l)\backslash j}(1-\hat\pi_{i}^{(0)})
        \over e^{-y} + (1-e^{-y})\prod\limits_{i\in N(l)\backslash j}
        (1-\hat\pi_{i}^{(0)})}}{
      1-(1-e^{-y})\hat\pi_j^{(0)} 
      {\prod\limits_{i\in N(l)\backslash j}(1-\hat\pi_{i}^{(0)})
        \over e^{-y} + (1-e^{-y})\prod\limits_{i\in{\cal N}(l)\backslash j}
        (1-\hat\pi_{i}^{(0)})}} \right\rangle \nonumber \\
  &=& {1 \over 2} \left\langle { 
      e^{-y} \sum\limits_{j\in {\cal N}(l)} \hat\pi_{j}^{(0)}
      \prod\limits_{i\in {\cal N}(l)\backslash j}(1-\hat\pi_{i}^{(0)})
      \over e^{-y} + (1-e^{-y}) \prod\limits_{i\in{\cal N}(l)} (1-\hat\pi_i^{(0)}) }
  \right\rangle \nonumber \\
  &=& {1\over 2} \langle \hat\pi_{l}^{(*)} \rangle 
\label{eq:energyconsist}
\end{eqnarray}

The mean complexity $\Sigma(c,x)$ can be calculated analogously,
cf.~Ref.~\cite{ZhouH2003} and Eqs.~(\ref{eq:Phi2},\ref{eq:Phi3}). The
final expression reads
\begin{eqnarray}
  \Sigma(c,x)&=&
  y x 
  + \sum\limits_{k=1}^{\infty} f_c(k) \prod\limits_{l=1}^{k}
  \int\limits_{0}^{1} {\rm d} \hat\pi_l^{(0)} P(\hat\pi_l^{(0)})
  \ln\bigl( e^{-y} + (1-e^{-y}) \prod\limits_{l=1}^{k} (1-\hat\pi_{l}^{(0)})
  \bigr) \nonumber \\
  & & -{c \over 2} \int\limits_{0}^{1} {\rm d} \hat\pi_1^{(0)} P(\hat\pi_{1}^{(0)})
  \int\limits_{0}^{1} {\rm d} \hat\pi_2^{(0)} P(\hat\pi_{2}^{(0)}) 
  \ln\bigl( 1-(1-e^{-y}) \hat\pi_1^{(0)} \hat\pi_2^{(0)} \bigr) \ .
  \label{eq:entropy}
\end{eqnarray}

\subsection{Optimal re-weighting and minimal VC density}

At given average degree $c$, Eq.~(\ref{eq:enew06}) allows to calculate
the typical VC size as a function of the re-weighting parameter $y$. It
is monotonously decreasing with growing $y$, so naively one would
expect that the minimal VC size is found in the limit of $y\to\infty$.
There is, however, a problem: The complexity $\Sigma(c,y)$ reaches
zero at some (a priori) $c$-dependent value $y^*$, and becomes negative
for larger $y$. Being defined as the logarithm of the number of
corresponding clusters, negative complexities correspond to VC sizes
typically non-existing in random graphs of mean degree
$c$. Consequently, we have to determine the size of the minimal VC of
a typical graph by $x(c, y^*)$ from the zero-complexity criterion
$\Sigma(c,y^*)=0$.

\begin{figure}[htb]
  \begin{center}
    \psfig{file=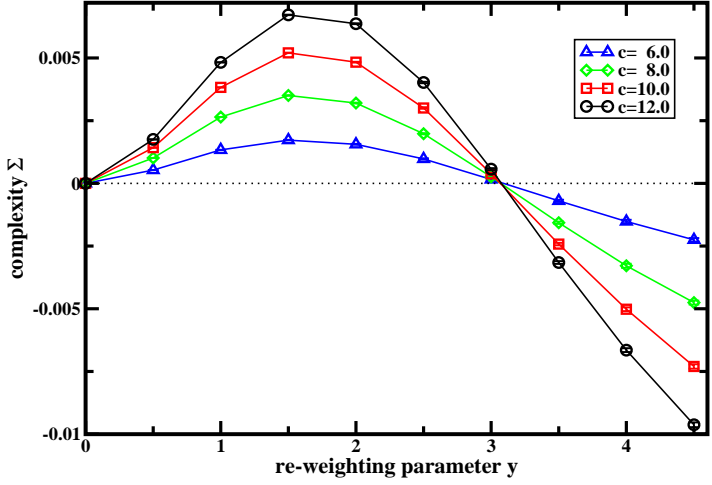,angle=270,width=10.0cm}
  \end{center}
  \caption{\label{fig:complexity} (Color online) Complexity $\Sigma(c,y)$ as a
    function of the re-weighting parameter $y$ for fixed mean vertex
    degree $c=12.0$ (circles), $c=10.0$ (squares), $c=8.0$ (diamonds),
    and $c=6.0$ (triangles).  All these curves seem to intersect in
    two points, $y=0$ and $y=y^*\approx 3.1$. In these two points,
    $\Sigma(c,y)=0$.}
\end{figure}

Figure~\ref{fig:complexity} shows $\Sigma(c,y)$ as a function of $y$
at various fixed $c$ values. At given $c$ value ($c>e$), the
complexity $\Sigma(c,y)$ first increases with $y$ as $y$ increases
from zero. $\Sigma(c,y)$ attains its maximal value when $y$ increases
to $y\approx 1.5$. Afterwords, $\Sigma(c,y)$ decreases with $y$ and it
reaches $\Sigma=0$ when $y=y^*\approx 3.1$. Upon further increase of
$y$, the complexity becomes negative. It is remarkable that the
$\Sigma(c,y)$ curves for different $c$ values intersect at (almost)
the same point $y^*$, which is just the point where the complexity
vanishes, $\Sigma(c,y^*)=0$. At present we do not understand why the
complexities for systems with different $c$ values should approaches
zero at (almost) the same point.

\begin{figure}[htb]
  \begin{center}
    \psfig{file=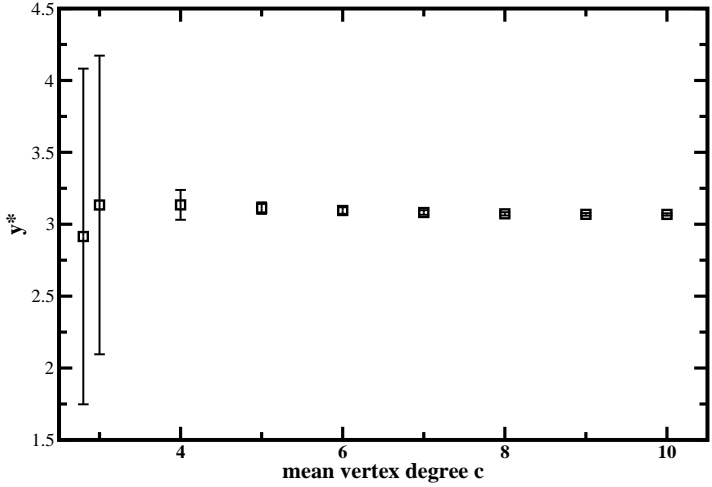,angle=270,width=10.0cm}
  \end{center}
    \caption{\label{fig:yyoptimal} Relationship between the
    optimal re-weighting parameter $y^*$ and the mean vertex degree
    $c$.  }
\end{figure}

At each fixed $c$ value, the optimal $y^*$ value can be determined
from the point of $\Sigma(c,y^*)=0$. The optimal $y^*$ value was
calculated numerically by population dynamics and shown in
Fig.~\ref{fig:yyoptimal} as a function of mean vertex degree $c$.
Figure~\ref{fig:yyoptimal} indeed demonstrates that the optimal
re-weighting parameter $y^*$ is insensitive to $c$ and stays at
$y^*\approx 3.1$ over the whole range of inspected $c$ values. This is
also in agreement with Ref.~\cite{ZhouH2003} (note that $y^*$ in the
present article corresponds to $2 y^*$ in \cite{ZhouH2003}). Even when
$c=2.8$ (just slightly beyond $e$) we have $y^*=2.9\pm 1.2$, which is
significantly different from zero, but consistent with a constant
$y^*$.  From Fig.~\ref{fig:yyoptimal} we thus get the impression that,
as the mean vertex degree $c$ exceeds $e$, the optimal re-weighting
parameter jumps quickly to a value $y^* \approx 3.1$.

\begin{figure}[htb]
  \begin{center}
    \psfig{file=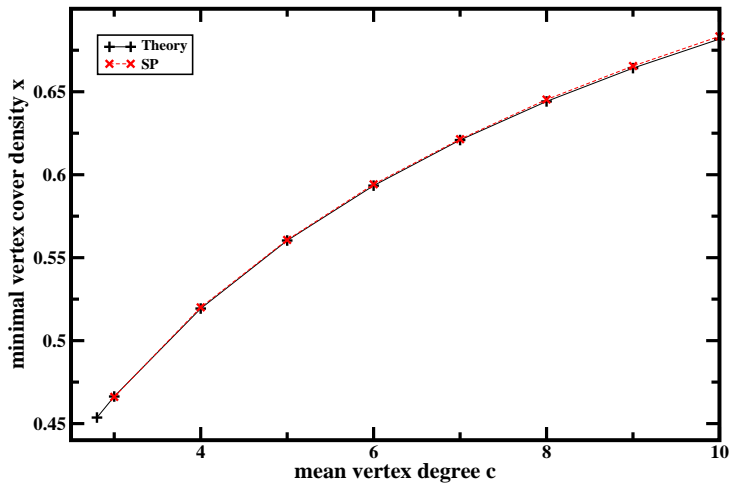,angle=270,width=10.0cm}
  \end{center}
  \caption{\label{fig:minimalvc} (Color online) The fraction of
    covered vertices (VC density) $x$ in a minimum vertex cover
    problem as a function of mean vertex degree $c$. Typical-case
    statistical physics results are given by a $+$, and $\times$ gives
    the estimates made by SP on graphs of size $N=5000$, averaged over
    20 samples.  }
\end{figure}

The minimum vertex cover size can also be obtained. In
Fig.~\ref{fig:minimalvc} we show the relationship between the minimal
vertex cover size and the mean vertex degree $c$.  As a comparison,
Fig.~\ref{fig:minimalvc} also includes the mean minimal vertex cover
size as estimated by the SP algorithm of the last section ($N=5000,\
y=2.0$, each point averaged over 20 samples). The results obtained by
SP and those obtained by the mean-field statistical physics
calculations are in very good agreement. At given vertex degree $c$,
the minimal VC density estimated by SP and the mean-field theory is
lower than the corresponding value obtained through exact enumeration
followed by extrapolation \cite{WeigtM2000}. The reason for such a
discrepancy can be understood.  According to
Fig.~\ref{fig:complexity}, at given mean vertex degree $c\leq 10$, the
maximum complexity of the system is less than $5\times 10^{-3}$. This
indicates that clustering of minimal VC solutions into distantly
separated domains will only occur for random graphs with size $N\geq
10^3$. For small random graphs as used in Ref.~\cite{WeigtM2000}, it
is very likely that all the minimal VC solutions can be grouped into a
single cluster (but with long-range frustrations among those vertices
described by the joker state $*$ \cite{ZhouH2005a}).

To summarize this subsection, we list in Table~\ref{tab:tabyyee} the
values of $y^*$ and the minimal energy density at several $c$
values. Theoretical and SP results are extremely close to each other,
even if the latter are systematically slightly larger. This may be due
to various reasons: SP uses finite size and really constructs a --
possibly non-optimal -- VC, whereas the theory works at the 1RSB level
which again is not exact due to higher RSB effects. We expect,
however, both estimates to be very close (but slightly different) from
the exact result.

\begin{table}
\caption{
  \label{tab:tabyyee}
  The optimal re-weighting parameter $y^*$ and the minimal VC density
  $x$ as estimated by the 1RSB ansatz and by SP ($N=5000$).  }
  \begin{tabular}{c|ccccccccc}
    $c$&$2.8$ &$3.0$ & $4.0$ & $5.0$ & $6.0$ & $7.0$ & $8.0$ & $9.0$ & $10.0$ 
    \\ \hline
    $y^*$ & $3(1)$ & $3(1)$ & $3.1(1)$ & $3.11(4)$ & $3.10(2)$ 
    & $3.08(2)$ & $3.07(1)$ & $3.069(9)$ & $3.068(8)$ \\ \hline
    $x$ (theory) & $0.4536290(8)$ & $0.46632(2)$ & $0.51934(2)$ 
    & $0.56033(3)$ & $0.59341(3)$ & $0.62088(2)$ & $0.64416(2)$
    & $0.66423(2)$ & $0.68175(2)$ \\ \hline
    $x$ (SP) & & 0.4661(5)& 0.52004(4) & 0.5607(3) & 0.5942(3) & 0.6214(3) 
    & 0.6453(2) & 0.6655(3) & 0.6834(2)  \\ \hline
  \end{tabular}
\end{table}

\subsection{Relaxation time of the population dynamics}
\label{subsec:relaxation}

Let us finally analyze the mean-field population dynamics which aims
at finding a fixed-point distribution for Eq.~(\ref{eq:pi00distr}). In
the population dynamics, an array of ${\cal N}$ values $\hat\pi^{(0)}$
is first initialized randomly (typically we use ${\cal N}\sim 10^{6}$,
this number should not be confused with the vertex number $N$ in the
single sample analysis of the previous sections). Then in each time
step, corresponding to an interval $\Delta t=1/{\cal N}$, we perform
the following update of the population:
\begin{enumerate}
\item[(1)] A natural number $k$ is drawn from the Poisson distribution
$f_c(k)$.
\item[(2)] $k$ elements $\hat\pi^{(0)}_i,\ i=1,...,k,$ are randomly and
independently chosen in the current population.
\item[(3)] A new $\hat\pi^{(0)}$ is calculated according to
Eq.~(\ref{eq:pi00}).
\item[(4)] A randomly selected element in the population is replaced with
this new $\hat\pi^{(0)}$ value.
\end{enumerate}
This iteration is repeated many times (typically of the order of
$10^{4} {\cal N}$), until the statistical properties of the population
approach stationary values. The histogram of the population is then
our estimate of the self-consistent distribution $P(\hat\pi_0)$ in
Eq.~(\ref{eq:pi00distr}).

Suppose that, at time $t$, the histogram of $\hat\pi^{(0)}$ over the
whole population is given by
\begin{equation}
  \label{eq:pi_t}
  P(\hat\pi^{(0)}; t) = p_1(t) \delta(\hat\pi^{(0)}) 
  + p_2(t) \delta(\hat\pi^{(0)}) 
  + p_3(t)  \rho(\hat\pi^{(0)};t) \ ,
\end{equation}
with $p_3(t)=1-p_1(t)-p_2(t)$, and with $\rho(\hat\pi^{(0)};t)$
satisfying the conditions
\begin{equation}
  \label{eq:rho}
  \rho(0;t)\equiv 0 \ , \;\;\;\; \rho(1; t) \equiv 0 \ , \;\;\;\; 
  \int_{0}^{1} {\rm d} \hat\pi^{(0)} \rho(\hat\pi^{(0)}; t) \equiv 1 \ .
\end{equation}
Since the population of $\hat\pi^{(0)}$ values at time $t+\Delta t$ is
obtained by replacing one randomly chosen element of the population of
time $t$ with the newly calculated $\hat\pi^{(0)}$, we can write down
the following two evolution equations for $p_1(t)$ and $p_2(t)$:
\begin{eqnarray}
  {\cal N} p_1(t+\Delta t)\ =\ {\cal N} p_1(t)
  + \sum\limits_{k=1}^{\infty} f_c(k) [1-\bigl(1-p_2(t)\bigr)^k]
  - p_1(t)  \ \ \ & \to & \ \ \  { {\rm d} p_1(t) \over {\rm d} t} 
  = 1 - e^{-c p_2(t)} - p_1(t) 
  \label{eq:p1t} \\
  {\cal N} p_2(t+\Delta t)\ =\ {\cal N} p_2(t) 
  + \sum\limits_{k=1}^{\infty}f_c(k) p_1^k - p_2(t) \hspace{2.63cm} 
  & \to & \ \ \
  { {\rm d} p_2(t) \over {\rm d} t} = e^{-c[1-p_1(t)]} - p_2(t) 
  \label{eq:p2t}
\end{eqnarray}
More precisely, these equations describe the average evolution over
many runs of the population dynamics. For large populations, ${\cal
N}\gg 1$, the true evolution of one population is, however, expected to
be closely concentrated around its expectation value, with random
fluctuations of the order ${\cal O}(1/\sqrt{{\cal N}})$. These equations
can be understood easily: In Eq.~(\ref{eq:p1t}), we describe the
expected number of zero-elements of the population. This number is
decreased by one with probability $p_1(t)$ by replacing an old zero
element, or it grows by one if a new zero element is introduced. The
latter case happens if in between the $k$ ``parents''
$\hat\pi^{(0)}_i,\ i=1,...,k,$ selected before, there exists at least
one which equals one. Analogously, a new element equal to one is
inserted in the population if all parents were equal to zero,
explaining the gain term in Eq.~(\ref{eq:p2t}).

The fixed-point solution $(p_1, p_2)$ of Eqs.~(\ref{eq:p1t}) and
(\ref{eq:p2t}) is determined by
\begin{equation}
  \label{eq:p12fix}
  p_1= 1-e^{-c p_2} \ , \;\;\;\; p_2=e^{-c (1-p_1)} \ .
\end{equation}
Note that for $c<e$, only one solution with $p_1+p_2=1$ exists. This
solution corresponds to replica-symmetry, only one solution cluster
exists, and consequently no cluster-to-cluster fluctuations
exist. Above $c=e$, also two other solutions with $p_1+p_2\neq 1$
exist. Only one is iterationally stable, it fulfills $p_1+p_2<1$ and
allows therefore for cluster-to-cluster variations of the value of
$\hat \pi^{(0)}_i$ for some vertices $i$.

To study the convergence velocity toward this fixed-point solution,
we assume that
\begin{equation}
  \label{eq:p12pert}
  p_1(t)=p_1+\epsilon_1(t) \ , \;\;\;\;
  p_2(t)=p_2+\epsilon_2(t) \ ,
\end{equation}
where $\epsilon_1(t)$ and $\epsilon_2(t)$ are small
quantities. Linearizing the dynamical equations, we find
\begin{equation}
  \label{eq:e12}
  {{\rm d}\epsilon_1(t) \over {\rm d} t}  
  = c(1-p_1) \epsilon_2(t)-\epsilon_1(t) \ , \;\;\;\;
  {{\rm d} \epsilon_2(t) \over {\rm d} t} 
  = c p_2 \epsilon_1(t)-\epsilon_2(t) \ ,
\end{equation}
and the typical relaxation time for $p_1(t)$ and $p_2(t)$ is
\begin{equation}
  \label{eq:tau12}
  \tau_{12}={1 \over 1- c \sqrt{(1-p_1) p_2}} \ .
\end{equation}
When the mean vertex degree $c$ approaches $e$ from below, 
the parameter $p_2$ approaches $e^{-1}$ as
\begin{equation}
  \label{eq:p2toem}
  p_2 \simeq {1\over e}+ {(e-c) \over 2 e^2}  \ .
\end{equation}
Consequently, the typical relaxation time $\tau_{12}$ diverges as
\begin{equation}
  \label{eq:tau12em}
  \tau_{12} \sim {2 e \over e - c } \, \;\;\;\;\;\;{\rm for}\ \ c < e \ .
\end{equation}
Note that the same critical behavior was found for the bug relaxation
time in the purely replica-symmetric warning-propagation equations.

On the other hand, when $c$ approaches $e$ from above, 
\begin{equation}
  \label{eq:p2toep}
  p_2 \simeq {1\over e} -\bigl({6 (c-e) \over e^3}\bigr)^{1/2}  
  + {(c-e) \over e^2} \ ;
\end{equation}
therefore $\tau_{12}$ diverges as
\begin{equation}
  \label{eq:tau12ep}
  \tau_{12} \sim { e \over c - e} \ , \ \ \ \ {\rm for} \ \ c > e \ .
\end{equation}
Equation (\ref{eq:tau12em}) and Eq.~(\ref{eq:tau12ep}) were confirmed
in single-graph message passing experiment. Note that the only region
where this convergence slows down is close to the replica-symmetry
breaking transition at $c=e$, where also the population dynamics slows
down critically. Note also, that this relaxation time does not depend
on the re-weighting parameter $y$.

We now study the evolution of $\rho(\hat\pi^{(0)};t)$ in
Eq.~(\ref{eq:pi_t}). For this purpose, in the population dynamics we
can set $p_1(t)\equiv p_1$, $p_2(t)\equiv p_2$, $p_3(t)=p_3 = 1-
p_1-p_2$ to their stationary values, and store only those
$\hat\pi^{(0)}$ values that satisfy $0 < \hat\pi^{(0)} < 1$ in the
population array. The distribution $\rho(\hat\pi^{(0)}; t+\Delta t)$
is related with $\rho(\hat\pi^{(0)}; t)$ by the following equation,
describing the expected number of population entries in the interval
$(\hat\pi^{(0)},\hat\pi^{(0)}+\Delta \hat\pi^{(0)})$:
\begin{eqnarray}
  {\cal N} \rho(\hat\pi^{(0)};t+\Delta t) \Delta \hat\pi^{(0)} &=&
  {\cal N} \rho(\hat\pi^{(0)};t) \Delta \hat\pi^{(0)} 
  -\rho(\hat\pi^{(0)};t) \Delta \hat\pi^{(0)} \nonumber \\
  & & + \sum\limits_{m=1}^{\infty} {f_{c p_3}(m) \over 1-e^{-c p_3}} 
  \prod\limits_{l=1}^{m} \int_{0}^{1} {\rm d} \hat\pi_{l}^{(0)} 
  \rho(\hat\pi_{l}^{(0)};t)  \delta\left( \hat\pi^{(0)} - 
  { \prod\limits_{l=1}^{m} (1-\hat\pi_{l}^{(0)}) \over e^{-y} 
    + (1-e^{-y}) \prod\limits_{l=1}^{m} (1-\hat\pi_{l}^{(0)}) } 
  \right) \Delta \hat\pi^{(0)}\ . \nonumber \\
  \label{eq:rho_discrete}
\end{eqnarray}
From Eq.~(\ref{eq:rho_discrete}) we see that
\begin{equation}
  \label{eq:rho_continue}
  {\partial \rho(\hat\pi^{(0)};t) \over \partial t} 
  = -\rho(\hat\pi^{(0)};t)
  + \sum\limits_{m=1}^{\infty} {f_{c p_3}(m) \over 1-e^{-c p_3}} 
  \prod\limits_{l=1}^{m} \int_{0}^{1} {\rm d} \hat\pi_{l}^{(0)} 
  \rho(\hat\pi_{l}^{(0)};t) 
  \delta\left( \hat\pi^{(0)} - 
  { \prod\limits_{l=1}^{m} (1-\hat\pi_{l}^{(0)}) \over 
    e^{-y} + (1-e^{-y}) \prod\limits_{l=1}^{m} (1-\hat\pi_{l}^{(0)}) } 
  \right) \ .
\end{equation}
The fixed-point solution of Eq.~(\ref{eq:rho_continue}) is
\begin{equation}
  \label{eq:rho_fp}
  \rho(\hat\pi^{(0)})
  =\sum\limits_{m=1}^{\infty} {f_{c p_3}(m) \over 1-e^{-c p_3} }
  \prod\limits_{l=1}^{m} \int {\rm d}  \hat\pi_l^{(0)} 
  \rho (\hat\pi_l^{(0)}) 
  \delta\Bigl( \hat\pi^{(0)}-{\prod\limits_{l=1}^{m} 
    (1-\hat\pi_{l}^{(0)}) \over e^{-y}
    + (1-e^{-y}) \prod\limits_{l=1}^{m}(1-\hat\pi_{l}^{(0)})} \Bigr) \ ,
\end{equation}
as can be seen also directly from Eqs.~(\ref{eq:pi00distr}) and
(\ref{eq:pi_t}).

Now let us suppose that, at time $t$, the actual distribution
$\rho(\hat\pi^{(0)};t)$ deviates from the fixed-point distribution
only slightly:
\begin{equation}
  \label{eq:rhosmall}
  \rho(\hat\pi^{(0)};t)=\rho(\hat\pi^{(0)})+\epsilon_3(\hat\pi^{(0)};t) \ ,
\end{equation}
with $|\epsilon_3(\hat\pi^{(0)};t)| \ll 1$ for all $0<\hat\pi^{(0)}<1$
and
\begin{equation}
  \label{eq:rhocondition}
  \epsilon_3(0;t)=\epsilon_3(1;t)\equiv 0 \ , \hspace{1.0cm}
  \int_{0}^{1} {\rm d} \hat\pi^{(0)} \epsilon_3(\hat\pi^{(0)};t) = 0 \ .
\end{equation}
The linearized evolution equation for $\epsilon_3(\hat\pi^{(0)};t)$ is
\begin{eqnarray}
  {\partial \epsilon_3(\hat\pi^{(0)};t) \over \partial t} 
  & = & -\epsilon_3(\hat\pi^{(0)};t) + c p_2 {e^y \over (1+(e^y-1) 
    \hat\pi^{(0)})^2}
  \epsilon_3\bigl({1-\hat\pi^{(0)}\over 1+(e^y-1)\hat\pi^{(0)}};t \bigr) 
  \nonumber \\
  & & +c p_3 \int_{0}^{1} {\rm d} \hat\pi_1^{(0)}
  \epsilon_3(\hat\pi_{1}^{(0)};t) \int_{0}^{1} {\rm d} \hat\pi_2^{(0)} 
  \rho(\hat\pi_2^{(0)}) 
  \delta\Bigl(\hat\pi^{(0)}-1+{\hat\pi_1^{(0)} (1-\hat\pi_2^{(0)}) 
    \over 1-(1-e^{-y})(1-\hat\pi_1^{(0)})(1-\hat\pi_{2}^{(0)})} \Bigr)  \ .
  \label{eq:epsilon3}
\end{eqnarray}

The stability of Eq.~(\ref{eq:epsilon3}) can be analyzed by
Fourier-expanding $\epsilon_3(\hat\pi^{(0)};t)$ in the following way:
\begin{equation}
  \label{eq:epsilon3_expand}
  \epsilon_3(\hat\pi^{(0)};t)= 
  \sum\limits_{m=1}^{\infty} a_m(t) \sqrt{2} \sin(\pi m \hat\pi^{(0)} ) \ ,
\end{equation}
with coefficients $a_m(t)$ satisfying the global constraint
\begin{equation}
  \label{eq:constraint}
  \sum\limits_{n=0}^{\infty} { a_{2 n + 1}(t) \over 2 n +1 } \equiv 0 \ .
\end{equation}

Based on Eq.~(\ref{eq:epsilon3}) and Eq.~(\ref{eq:epsilon3_expand}),
one can write down the evolution equation for $a_m(t)$:
\begin{equation}
  \label{eq:amt}
  {d a_m(t) \over d t} = \sum\limits_{n=1}^{\infty} 
  \Lambda_{m n} a_n(t) \ ,
\end{equation}
where the elements of the matrix $\Lambda$ can be easily written down.

The task now is to identify the dominant eigen mode of
Eq.~(\ref{eq:amt}) under the constraint of
Eq.~({\ref{eq:constraint}). We have performed such an analysis for
various $c$ values in the range $3 \leq c \leq 30$ and various $y$
values ranging from $y=0$ to $y=5$. In all the cases studied, the
dominant eigen mode of Eq.~(\ref{eq:amt}) decays to zero very quickly,
indicating that the mean-field population dynamics is exponentially
fast converging toward its fixed point. Compared to the iterative
stability of SP on single instances of random graphs, we find that the
messages may converge in population even if they do not converge on
the single graph any more. This is interesting since it allows to
extend the typical-case estimates to a region, where SP applied to
single samples fails to predict anything.

\section{Conclusion}
\label{sec:conclusion}

In this paper, we have formulated two message passing procedures for
solving -- or approximating -- the minimal vertex cover problem,
namely warning propagation and survey propagation. We have analyzed
the performance of both algorithms on the test bed of
finite-connectivity random graphs, where previous statistical-physics
approaches based both on the replica approach and on the cavity method
provide an insight on the phase diagram. We have also discussed in
detail how the message-passing approach is technically connected to
these typical-case based statistical-physics results.

For small average vertex degrees $c<e$ replica symmetry is known to
hold in the space of all minimal vertex covers. Therefore the simpler
one of the two algorithms -- warning propagation, which is based on
the replica-symmetric Bethe-Peierls iterative approach -- is
applicable. Comparing it to the exact leaf-removal algorithm, we have
shown that it outputs true minimal vertex covers. Unfortunately the
iterative solution of the warning propagation equations slows down
critically if we approach $c=e$ from below, and it does not converge
at all for higher average degrees. 

In this higher-degree region, replica symmetry is known to be
broken. We have therefore applied a survey propagation algorithm which
is based on the first step of replica symmetry breaking. We have
identified a parameter range where the message passing equations
converge to a globally self-consistent solution, which can be used to
construct small vertex covers. We have found that the provided results
not only outperform simple local-search procedures, but are consistent
to exact asymptotic results for high but finite average
degrees. Interestingly the vertex covers produced at the end are
relatively insensitive to details of the algorithms (in particular to
the somewhat heuristic choice of the re-weighting parameter $y$).

In the case of vertex cover, replica symmetry is known to be fully
broken, i.e. the exact solution for $c>e$ is known to be more
complicated than the one described by one-step replica-symmetry
breaking. Intuitively the solutions are expected to be organized in
clusters of clusters of clusters etc. So, even if the results of the
application of survey propagation are very promising, it is expected
to provide only a good approximation algorithm. It could therefore be
interesting to go beyond survey propagation and to formulate an
algorithm based on the second step of replica symmetry breaking
(corresponding to two hierarchical clustering levels), in order to see
if the higher complexity of the algorithm required is leading to even
smaller vertex covers than survey propagation.

As a last point, it could be interesting to apply the algorithm to
real-world covering problems, possibly extending it to the specific
nature of these tasks, which may be similar but not equal to the
original minimal VC problem \cite{Va,Gomez}. These problems are
frequently characterized by a broad degree distribution of the
underlying networks. Their extreme heterogeneity may result in a
better performance of simple heuristic algorithms exploiting local
network structures. On the other hand, it was shown in \cite{VaWe}
that assortative degree correlations may force replica-symmetry to
break also in scale-free networks, and algorithms like survey
propagation are expected to become efficient. A related interesting
question is the local network structure beyond the vertex degrees, in
particular small loops or other small dense subgraphs. It may be
practically necessary to coarse grain the graph considering such loops
as single constraints, and by applying the factorization hypothesis
only to larger structures. In the replica symmetric case (warning or
belief propagation), this corresponds to the region-graph method
proposed in \cite{YeFrWe}, for the one-step replica symmetry broken
case (survey propagation) it is still an open technical challenge.

{\bf Acknowledgments:} We gratefully acknowledge discussions
W. Barthel, A.K. Hartmann and G. Semerjian.

\bibliography{vc_sp}
\end{document}